\documentclass[aps,twocolumn,superscriptaddress,longbibliography,amsmath,amssymb,amsfonts,citeautoscript]{revtex4-1}
\usepackage{graphicx}
\usepackage{bm}
\usepackage{bbm}
\usepackage{color}
\usepackage{epstopdf}
\usepackage{amsmath}
\usepackage{amssymb}
\usepackage{ulem}
\usepackage[urlcolor=blue,colorlinks=true,citecolor=blue,linkcolor=blue,pdfstartview={FitH},bookmarks=false]{hyperref}
\usepackage{todonotes}
\usepackage{ulem}
\usepackage{cancel}
\usepackage{verbatim}
\usepackage[T1]{fontenc}

\newcommand{\TD}[1]{\textcolor{black}{#1}}
\begin{document}

\title{Transient triplet blockade in Andreev junction}

\author{R. Taranko}
\affiliation{Institute of Physics, M. Curie-Sk\l{}odowska University, 20-031 Lublin, Poland}

\author{J. Bara\'nski}
\affiliation{Department of General Education, Polish Air Force University, ul. Dywizjonu 303 nr 35, 08521 D\k{e}blin, Poland}             

\author{A. Jankiewicz}
\affiliation{Institute of Spintronics and Quantum Information, Faculty of Physics and Astronomy, A. Mickiewicz University, 61-614 Pozna\'n, Poland}

\author{K. Wrze\'sniewski}
\affiliation{Institute of Spintronics and Quantum Information, Faculty of Physics and Astronomy, A. Mickiewicz University, 61-614 Pozna\'n, Poland}

\author{I. Weymann}
\affiliation{Institute of Spintronics and Quantum Information, Faculty of Physics and Astronomy, A. Mickiewicz University, 61-614 Pozna\'n, Poland}

\author{T. Doma\'nski}
\affiliation{Institute of Physics, M. Curie-Sk\l{}odowska University,
             20-031 Lublin, Poland}

\date{\today}

\begin{abstract}
We study the time-dependent triplet configuration, appearing under nonequilibrium conditions in a nanoscopic junction with two quantum dots coupled in series between superconductor and normal metallic lead. We show that in the situation, when both quantum dots are singly occupied by identical spin electrons, the on-dot electron pairing is suppressed what substantially affects the subgap charge transport. We investigate processes in which such configuration can be temporarily encountered, either due to the initial conditions or by imposing the external magnetic field. 
Our analytical and numerical calculations provide estimations for the temporal scales, characterizing evolution of the triplet configuration which could be manifested in the time-resolved tunneling measurements. Such nonequilibrium features of the triplet configuration might be relevant to operations on superconducting qubits, in their conventional and/or topological realizations. 
\end{abstract}

\maketitle

\section{Introduction}
\label{sec.intro}

Double quantum dots embedded in various tunneling structures can be a platform for the realization of quantum bits (qubits) and the implementation of quantum information processing \cite{Kouwenhoven-2002,Nowack-2007}. Specifically, quantum dots confined in superconducting hybrid structures are currently used for constructing the superconducting qubits with long coherence times and flexible control \cite{Krantz-2019,Aguado-2020}. Further perspectives are related with the topologically nontrivial states which can be engineered  in quantum dot ensembles coupled to superconductors \cite{SeoaneSouto2024,Prada-2020}, where the Majorana-based parity qubits could be realized.

Typical means to read/write qubits rely on charge sensing \cite{charge_sensing-2024,Aasen-2016} that depends on  electronic configuration of the quantum dots controllable by side-gate potentials. For instance, applying a weak voltage (smaller than the pairing gap) to the setup comprising the quantum dot placed between  superconductor and normal metallic lead can induce the subgap current by the  electron-to-hole scattering. Efficiency of this Andreev mechanism is very sensitive to a specific configuartion of the in-gap bound states driven by the superconducting proximity effect \cite{balatsky.vekhter.06,Rodero-11}. We shall consider such bound states both in the unpolarized quantum dots (Sec.\ \ref{transient.blockade}) which have been initially predicted by Machida and Shibata \cite{Machida-1972}, and in the magnetically polarized dots  (Sec.\ \ref{Zeeman.blockade}), resembling Yu-Shiba-Rusinov states of classical magnetic impurities \cite{balatsky.vekhter.06}.

In superconducting structures with two quantum dots, the Andreev current can be blocked if both dots are singly occupied by the same spin electrons. This phenomenon, dubbed `Andreev blockade', has been predicted \cite{Pekker-2021} and indeed observed experimentally \cite{Zhang-2022}. The `triplet blockade' has been also detected, using two quantum dots placed in series in the Josephson junction \cite{Paaske-2020,Souto-2023} and similar phenomena might be encountered in the topologically nontrivial superconducting structures \cite{Flensberg-2024,spin_filtered_Andreev-2023}.

\begin{figure}
\includegraphics[width=0.95\linewidth]{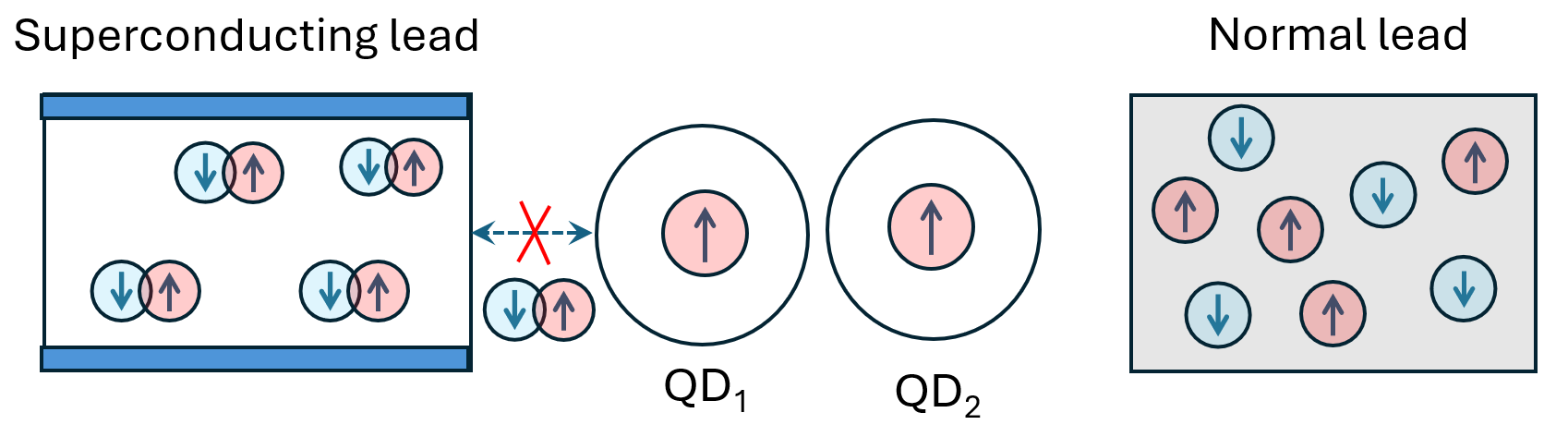}
\caption{Schematics of the spin triplet configuration in two quantum dots (QD$_{1,2}$) coupled between the normal (N) and superconducting (S) leads, which blocks a transfer of the Cooper pairs on QD$_{1}$.}
\label{fig.scheme}
\end{figure}  

In this work we explore time-dependent realizations of the triplet configuration appearing under nonequilibrium conditions in the setup displayed in Fig.~\ref{fig.scheme}. Our study shows that temporary existence of the triplet configuration leads to suppression of the electron pairing in both quantum dots, affecting the subgap charge transport. Details of these time-resolved  phenomena are considered by us in three protocols.
\begin{itemize}
\item{We analyze evolution of the triplet configuration settled by the initial conditions and determine the characteristic time-scale(s) needed for charge redistribution, lifting the triplet state and enabling a buildup of the unpolarized in-gap bound states of the strongly/weakly hybridized dots.}
\item{We investigate gradual development of the triplet configuration induced by applying the external magnetic field and show its detrimental influence on the electron pairing in both quantum dots. We also show that it suppresses the subgap current in the infinitesimally biased junction, in close analogy to the Andreev blockade discussed in Refs. \cite{Pekker-2021,Zhang-2022} for the correlated setup.}
\item{
Going beyond the linear response limit, we reveal that sufficiently strong bias voltage applied across the junction is capable to eliminate the triplet configuration, leading to a recovery of the Andreev transport in presence of arbitrary magnetic field.}
\item{
We briefly address variation of the electron pairing and the inter-dot spin correlations induced by the external magnetic field of the interacting quantum dots. We find that rearrangement of the bound states energies due to the magnetic field can be manifested by nonmonotonic behavior, indicating a subtle interplay of the Coulomb repulsion with the superconducting proximity effect.}
\end{itemize}

Our study is complementary to previous investigations of the subgap charge transport under static conditions which has been analyzed for different realizations of the double quantum dots by various methods \cite{Choi-2000,Zhu-2002,Tanaka.2010,Zitko.2010,
Konig.2010,Droste.2012,Grifoni.2013,Brunetti-2013,Yao2014Dec,Sothmann-2014,Trocha2015Jun,
Meng.2015,Zitko-2015,Su.2017,Wrzesniewski2017Nov,Glodzik.2017,Scherubl_2019,
Wojcik2019Jan,Wang-2019,Leijnse.2019,Franke-2022,Zalom2024Oct}. Numerous experimental data have been obtained for the semiconducting nanowires \cite{Sherman.2017,Su.2017, Grove_Rasmussen.2018,Estrada_Saldana.2018,Estrada_Saldana.2020,
Paaske-2020,Zhang-2022} and carbon nanotubes \cite{Cleuziou.2006,Pillet.2013} 
hybridized with superconducting lead(s), and  for dimers  on surfaces of superconductors 
\cite{Ruby.2018,Franke-2018,Choi.2018,Kezilebieke.2019,Franke-2024}. 
We focus on evolution of the triplet configuration and analyze its influence on the pairing amplitudes, affecting the subgap charge transport. For this purpose, we extend the formalism introduced in Ref.~\cite{Taranko-2021} to the spin-dependent energy levels what gives us insight into the dynamics of the triplet configuration. 

The paper is structured as follows. In Sec.\ \ref{sec.model} we introduce the microscopic model. Next, in Sec.~\ref{transient.blockade}, we discuss the transient effects after forming S-QD$_{1}$-QD$_{2}$-N junctions with the initial triplet configuration of quantum dots. Section~\ref{Zeeman.blockade} describes development of the triplet configuration induced by the external magnetic field and we also analyze its partial/complete lifting by a source-drain voltage. Technical details concerning the time-dependent observables are presented in Appendices \ref{app_Laplace}-\ref{app_Pairing_chi11}. Selected numerical results obtained by NRG calculations for the correlated system are presented in Appendix \ref{correlation_effects}.

\section{Microscopic model}
\label{sec.model}

Our setup, comprising two quantum dots (QD$_{1,2}$) coupled in series between the superconducting (S) and the normal (N) leads, can be described by the Hamiltonian
\begin{equation}
\hat{H}=\hat{H}_{S}+\hat{H}_{N}+\hat{H}_{hybr}+\sum_{j=1,2} \hat{H}_{{QD}_{j}} .
\label{Hamil_N_DQD_S}
\end{equation}
We treat the normal metallic lead as a free fermion gas
\begin{equation}
\hat{H}_{N}=\sum_{ {\bf{k}}\sigma}\xi_{N {\textbf{k}}}
\hat{c}_{N {\textbf{k}}\sigma}^{\dagger}\hat{c} _{N {\textbf{k}}\sigma} ,
\end{equation}
where the energies $\xi_{N {\textbf{k}}}=\varepsilon_{N {\textbf{k}}}-\mu_{N}$  are measured with respect to the chemical potential $\mu_{N}$. The superconducting electrode is described by the BCS-type model
\begin{equation}
\hat{H}_{S}=\sum _{ {\textbf{q}}\sigma}\xi_{S {\textbf{q}}}\hat{c} _{S {\textbf{q}}\sigma}^{\dagger}\hat{c} _{S {\textbf{q}}\sigma}
-\sum _{ {\textbf{q}}}\left( \Delta_{sc} \hat{c}_{S {\textbf{q}}\uparrow}^{\dagger}\hat{c}_{S- {\textbf{q}}\downarrow}^{\dagger}+ \mbox{\rm h.c.} \right),
\label{eq:1}
\end{equation}
assuming isotropic pairing gap $\Delta_{sc}$. The operators $\hat{c}_{\beta {\textbf{k}}\sigma}^{(\dagger)}$ refer to the annihilation (creation) of electrons whose energies $\xi_{S {\textbf{q}}}=\varepsilon_{S {\textbf{q}}}-\mu_{S}$ are expressed with respect to the chemical potential. In the presence of the bias voltage $V$ applied across the junction, the chemical potentials are detuned, $\mu_{N}-\mu_{S}=eV$. For convenience, we assume the superconducting lead to be grounded, $\mu_{S}=0$.

The quantum dots embedded between $N$ and $S$ leads will be treated as the Anderson-type impurities
\begin{eqnarray}
\hat{H}_{{QD}_{j}}=\sum_{\sigma}\varepsilon_{j\sigma}\hat{c}_{j\sigma}^{\dagger}\hat{c}_{j\sigma}
+U_j\hat{c}_{j\uparrow}^{\dagger}\hat{c}_{j\uparrow}\hat{c}_{j\downarrow}^{\dagger}\hat{c}_{j\downarrow}
\end{eqnarray}
and (in presence of the external magnetic field) their energies $\varepsilon_{j\sigma}$ can be spin-dependent.  $U_j$ is the potential of the repulsive Coulomb interaction between opposite spin electrons 
and the last part of Eqn.\ (\ref{Hamil_N_DQD_S}) describes the hybridization of the quantum dots with external leads 
\begin{eqnarray}
\hat{H}_{hybr} &=& \sum_{\sigma} \left(  
\sum_{{\textbf{q}}}V_{S {\textbf{q}}} \hat{c}_{S {\textbf{q}}\sigma}^{\dagger}\hat{c}_{1\sigma}
+ V_{12}\hat{c}_{1\sigma}^{\dagger}\hat{c}_{2\sigma} 
\right. \nonumber \\
&+& \left. 
\sum_{{\textbf{k}}}V_{N {\textbf{k}}} \hat{c}_{N {\textbf{k}}\sigma}^{\dagger}\hat{c}_{2\sigma} \right) + {\mbox{\rm h.c.}} \,.
\label{eq: 2}
\end{eqnarray}
Here $V_{12}$ is the inter-dot coupling  and $V_{\beta {\textbf{k}}}$ refers to the coupling of QD$_{\beta}$ with the neighboring lead, respectively. 

Focusing on the subgap charge transport, we assume the large bandwidth approximation and introduce the constant (energy-independent)  coupling $\Gamma_{\beta}=2\pi\sum_{ {\textbf{k}}} |V_{\beta {\textbf{k}}}|^{2} \delta(\omega - \varepsilon_{\beta {\textbf{k}}})$.
We shall study dynamical phenomena in the subgap region, treating the pairing gap $\Delta_{sc}$ as the largest energy scale. 
Such superconducting atomic limit approach, $\Delta_{sc}\to\infty$, can be regarded as an approximation that captures the main low-energy properties of QD$_1$-S hybrid part. For finite values of $\Delta_{sc}$, the following effects might arise: (i) slight renormalization of the Andreev bound state energies \cite{Bauer-2007,Baraski2013}, and (ii) additional damping effects driven by the quasiparticle states from outside the pairing gap of superconductor \cite{Taranko-2021}. In typical experimental realizations using Al ($\Delta_{sc} \approx 0.18$~meV) or Pb ($\Delta \approx 1.1$~meV), the pairing gap is one order of magnitude larger than the reported hybridization strength  ($\Gamma_S \sim 100$--$200~\mu$eV), see for instance Fig.~3 in Ref.\ \cite{Wernsdorfer-2012}. Under such conditions, the bound states are well separated from the gap edges and damping by the quasiparticles from outside the pairing gap of superconductor would not affect significantly our results (at least qualitatively) because the main source of the relaxation processes are continuum states of the normal metallic lead.

In the superconducting limit approach the influence of  the bulk superconductor on  QD$_{1}$ can be described by the following term
\begin{equation}
\hat{H}_{S}+
\sum_{\sigma,{\textbf{q}}} \left( V_{S {\textbf{q}}} 
\hat{c}_{S {\textbf{q}}\sigma}^{\dagger}\hat{c}_{1\sigma}
+\mbox{\rm h.c.}
\right)
\overset{\Delta_{sc}\rightarrow\infty}{\approx} \frac{\Gamma_{S}}{2}(\hat{c}^{\dagger}_{1 \uparrow} \hat{c}^{\dagger}_{1 \downarrow} +\hat{c}_{1 \downarrow}\hat{c}_{1 \uparrow}) ,
\end{equation}
where $\frac{\Gamma_{S}}{2}\hat{c}^{\dagger}_{1 \uparrow} \hat{c}^{\dagger}_{1 \downarrow}$ is the source and $\frac{\Gamma_{S}}{2} \hat{c}_{1 \downarrow}\hat{c}_{1 \uparrow}$ the sink terms of the proximity induced electron pair, respectively.

To simplify our notation, we set $\hbar=e=k_{B}=1$ and use the coupling $\Gamma_{S}\equiv 1$ as a convenient unit for all energies.  
For clarity, however, in all figures we explicitly remark that time is expressed in units of $\hbar/\Gamma_S$.

\section{Effects due to initial conditions}
\label{transient.blockade}

We start by studying \TD{the triplet configuration} after coupling the constituents of our system (at $t=0^{+}$) in the following initial state
\begin{eqnarray}
n_{j\sigma}(t \leq 0) =
\left\{ \begin{array}{ll}
1 & \hspace{0.3cm} \mbox{\rm for } \sigma=\uparrow , \\
0 & \hspace{0.3cm} \mbox{\rm for } \sigma=\downarrow  ,
\end{array} \right.
\label{QD_occup_def}
\end{eqnarray}
where $n_{j\sigma}$ stands for the spin-dependent occupancy of  $j$-th dot. We investigate the quantum evolution of our system (for $t>0$), neglecting the correlation effects, $U_j=0$. Under such circumstances, the time-dependent occupancy of quantum dots, $n_{j\sigma}(t)=\langle\hat{c}_{j\sigma}^{\dagger}(t)\hat{c}_{j\sigma}(t)\rangle$, on-dot pairing amplitudes, $\chi_{jj}(t)=\langle\hat{c}_{j\downarrow}(t)\hat{c}_{j\uparrow}(t)\rangle$, and other observables can be determined analytically. Their evolution from the initial configuration (\ref{QD_occup_def}) occurs via the transient currents. We inspect them for selected $\varepsilon_{j\sigma}$ and $V_{12}$, assuming the asymmetric coupling $\Gamma_{N}=0.2\Gamma_{S}$.

Let us outline the formalism used for such analysis based on our previous studies \cite{Taranko-2021,Taranko-2018}. The expectation value $\langle \hat{O}\rangle$ of an arbitrary observable $\hat{O}$ can be obtained from the Heisenberg equation of motion ${i \frac{d}{dt}\hat{O} =\big[ \hat{O},\hat{H}\big]}$. Here of primary importance are the time-dependent operators $\hat{c}_{j\sigma}^{(\dagger)}(t)$. We determine them from the coupled equations of motion, taking into account the initial conditions within the Laplace transforms ${\int^{\infty}_{0} dt e^{-st}\hat{c}_{j\sigma}^{(\dagger)}(t)=\cal{L}} \left\{ \hat{c}_{j\sigma}^{(\dagger)}(t) \right\}(s)\equiv \hat{c}_{j\sigma}^{(\dagger)}(s)$. Explicit expressions of these transforms $\hat{c}_{j\sigma}(s)$ for arbitrary $\varepsilon_{j\sigma}$ are presented in Appendix \ref{app_Laplace}. The time-dependent operators are obtained in the next step from their inverse Laplace transforms $\hat{c}_{j\sigma}^{(\dagger)}(t)={\cal{L}}^{-1}\left\{ \hat{c}_{j\sigma}^{(\dagger)}(s) \right\}(t)$.

As an example, the time-dependent population 
of spin-$\uparrow$ electrons of QD$_{1}$ computed from 
the general formula $n_{i\sigma}(t)=\left< {\cal{L}}^{-1} \left\{ \hat{c}_{i\sigma}^{\dagger}(s) \right\}(t) {\cal{L}}^{-1} \left\{ \hat{c}_{i\sigma}(s) \right\}(t)\right>$ is given by
\onecolumngrid
\begin{eqnarray}
&& n_{1\uparrow}(t) = n_{1\uparrow}(0)
\left| {\cal{L}}^{-1} \left\{ \frac{(s+i\varepsilon_{2\uparrow}+g)A_{1}(s)}
{A_{3}(s)}\right\} (t) \right|^{2} 
+n_{2\uparrow}(0) V_{12}^{2}
\left| {\cal{L}}^{-1} \left\{ \frac{A_{1}(s)}
{A_{3}(s)}\right\} (t) \right|^{2} \label{n_up_transient}  \\
&&+\left[ 1\!-\!n_{1\downarrow}(0)\right] \Delta^{2}
\left| {\cal{L}}^{-1} \left\{ \frac{(s+i\varepsilon_{2\uparrow}+g)(s-i\varepsilon_{2\downarrow}+g)}{A_{3}(s)}\right\} (t) \right|^{2} 
+\left[ 1\!-\!n_{2\downarrow}(0)\right] \Delta^{2} V_{12}^{2}
\left| {\cal{L}}^{-1} \left\{ \frac{(s+i\varepsilon_{2\uparrow}+g)}
{A_{3}(s)}\right\} (t) \right|^{2} \nonumber \\
&&+\frac{\Gamma_{N}}{2\pi} \Delta^{2}V_{12}^{2}
\int_{-\infty}^{+\infty} d\varepsilon \left[ 1\!-\!f_{N}(\varepsilon)\right] \left|
{\cal{L}}^{-1} \left\{ \frac{(s+i\varepsilon_{2\uparrow}+g)}
{(s-i\varepsilon)A_{3}(s)}\right\} (t) \right|^{2}   
+  \frac{\Gamma_{N}}{2\pi} V_{12}^{2}
\int_{-\infty}^{+\infty} d\varepsilon f_{N}(\varepsilon)  
\left|
{\cal{L}}^{-1} \left\{ \frac{A_{1}(s)}
{(s+i\varepsilon)A_{3}(s)}\right\} (t) \right|^{2} ,
\nonumber
\end{eqnarray}
\twocolumngrid
\noindent
where $f_{N}(\varepsilon)=\left[1+\mbox{\rm exp}\left((\varepsilon-\mu_{N})/k_{B}T\right)\right]^{-1}$ is the Fermi-Dirac distribution function. Additionally  we introduced the following  abbreviations $g=\frac{\Gamma_N}{2}$, $\Delta=\frac{\Gamma_{S}}{2}$, and 
\begin{eqnarray}
A_{1}(s)&=&\left( s-i\varepsilon_{1\downarrow}\right) 
\left( s -i\varepsilon_{2\downarrow}+g\right) + V_{12}^{2} ,
\label{Aa_1}
\\
A_{2}(s)&=&\left( s+i\varepsilon_{1\uparrow}\right) 
\left( s +i\varepsilon_{2\uparrow}+g\right) + V_{12}^{2} ,
\label{Aa_2}
\\
A_{3}(s)&=&\Delta^{2}
\left( s-i\varepsilon_{2\downarrow}+g\right) 
\left( s +i\varepsilon_{2\uparrow}+g\right)    
\nonumber \\ &+& A_{1}(s)A_{2}(s) .
\label{Aa_3}
\end{eqnarray}
The first four terms of Eq.~(\ref{n_up_transient}) describe the quantum oscillations which (for $\Gamma_{N}\neq 0$) vanish in the asymptotic limit, $t\rightarrow\infty$. The other terms (with the Fermi-Dirac function) represent dissipative parts, originating from a continuum of the normal metallic lead. Similar expressions for $n_{1\downarrow}(t)$ and $n_{2\sigma}(t)$ are presented in Appendix \ref{app_Laplace}.

Redistribution of the charge occupancies $n_{j\sigma}(t)$ from their initial values is realized via the transient currents $j_{N\sigma}(t)$, $j_{12\sigma}(t)$ which are sensitive to the electron pairing amplitudes $\chi_{jj}(t)$. The charge current flowing from the normal lead to QD$_2$ 
$j_{N\sigma}(t)= \left< \frac{d}{dt} \sum_{\bf k} \hat{c}^{\dagger}_{N{\bf k}\sigma}(t) \hat{c}_{N{\bf k}\sigma}(t)\right>
=2\mbox{\rm Im} \sum_{\bf k} V_{N{\bf k}} \left< \hat{c}_{2\sigma}^{\dagger}(t) \hat{c}_{N{\bf k}\sigma}(t) \right>$ can be expressed in the wide-bandwidth limit as \cite{Taranko-2018}
\begin{equation}
j_{N\sigma}(t)=2\mbox{\rm Im} \sum_{\bf k} V_{N{\bf k}} e^{-i\xi_{N{\bf k}}t} \left< \hat{c}_{2\sigma}^{\dagger}(t) \hat{c}_{N{\bf k}\sigma}(0) \right>-\Gamma_{N} n_{2\sigma}(t) .
\label{eqn_12}
\end{equation}
Using the Laplace transform $\hat{c}_{2\sigma}(s)$ [see Appendix~\ref{app_Laplace}] we can recast Eqn.\ (\ref{eqn_12}) into the following form
\begin{eqnarray}
&&j_{N\uparrow}(t) = \frac{\Gamma_N}{\pi} \mbox{\rm Re} \int_{-\infty}^{\infty} d\varepsilon f_{N}(\varepsilon) e^{-i\varepsilon t} \label{j_Nup} \\
&\times &
{\cal{L}}^{-1} \left\{ \frac{A^{*}_{3}(s)-V_{12}^{2}A^{*}_{1}(s)}{(s-i\varepsilon_{2\uparrow}+g)(s-i\varepsilon)A^{*}_{3}(s)}\right\} (t)-\Gamma_N n_{2 \uparrow}(t) .
\nonumber
\end{eqnarray}
In similar way, we can determine the inter-dot current $j_{12\sigma}(t)=-2V_{12} \mbox{\rm Im} \left< \hat{c}_{1\sigma}^{\dagger}(t) \hat{c}_{2\sigma}(t) \right>$.
The charge flowing from the superconductor to QD$_{1}$ is contributed by both spin components, ${j_S(t)=\sum_\sigma j_{S\sigma}(t)}$, where $j_{S\sigma}(t) =2\mbox{\rm Im} \sum_{\bf q} V_{S{\bf q}} \left< \hat{c}_{1\sigma}^{\dagger}(t) \hat{c}_{S{\bf q}\sigma}(t) \right>$ obeys the conservation law
\begin{eqnarray}
j_{S\sigma}(t)= j_{12\sigma}(t) + \frac{\partial}{\partial t} n_{1\sigma}(t).
\label{eqn.14}
\end{eqnarray}
In the superconducting atomic limit, the net current is equivalently expressed by $j_{S}(t)=-2\Gamma_S {\rm Im}\langle c_{1 \downarrow}(t)c_{1 \uparrow}(t)\rangle$ (see Appendix~\ref{app_Laplace} for details) and depends on the time-dependent pairing amplitude
\begin{eqnarray}
    \chi_{11}(t)
    =\left<{\cal{L}}^{-1} \left\{ c_{i \downarrow}(s)\right\}(t){\cal{L}}^{-1} \left\{ c_{i \uparrow}(s)\right\}(t)\right> 
\end{eqnarray}
whose explicit form is presented in Appendix \ref{app_Laplace}.

In what follows, we discuss numerical results obtained for the strongly and weakly coupled quantum dots. In this section we performed our numerical computations for the unbiased junction, therefore all transient currents vanish in the asymptotic limit, $t\rightarrow \infty$. 

\subsection{Strong inter-dot coupling}

Two strongly hybridized quantum dots form a dimer, mutually sharing their in-gap bound states. The strong coupling, $V_{12}$, ensures that a buildup of these {\it molecular} in-gap states is expected to occur nearly simultaneously.

\begin{figure}[b!]
\includegraphics[width=0.95\linewidth]{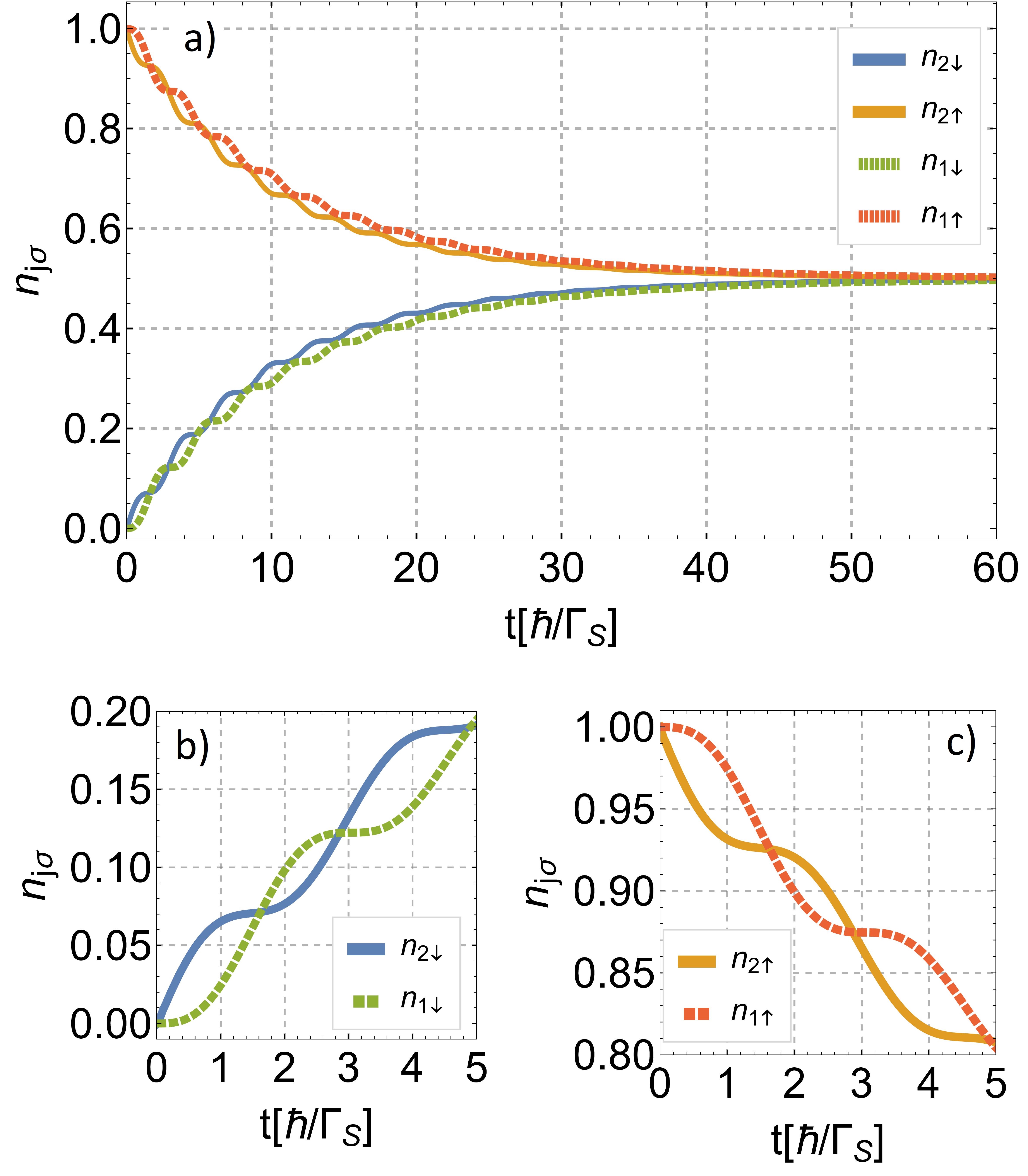}
\includegraphics[width=0.95\linewidth]{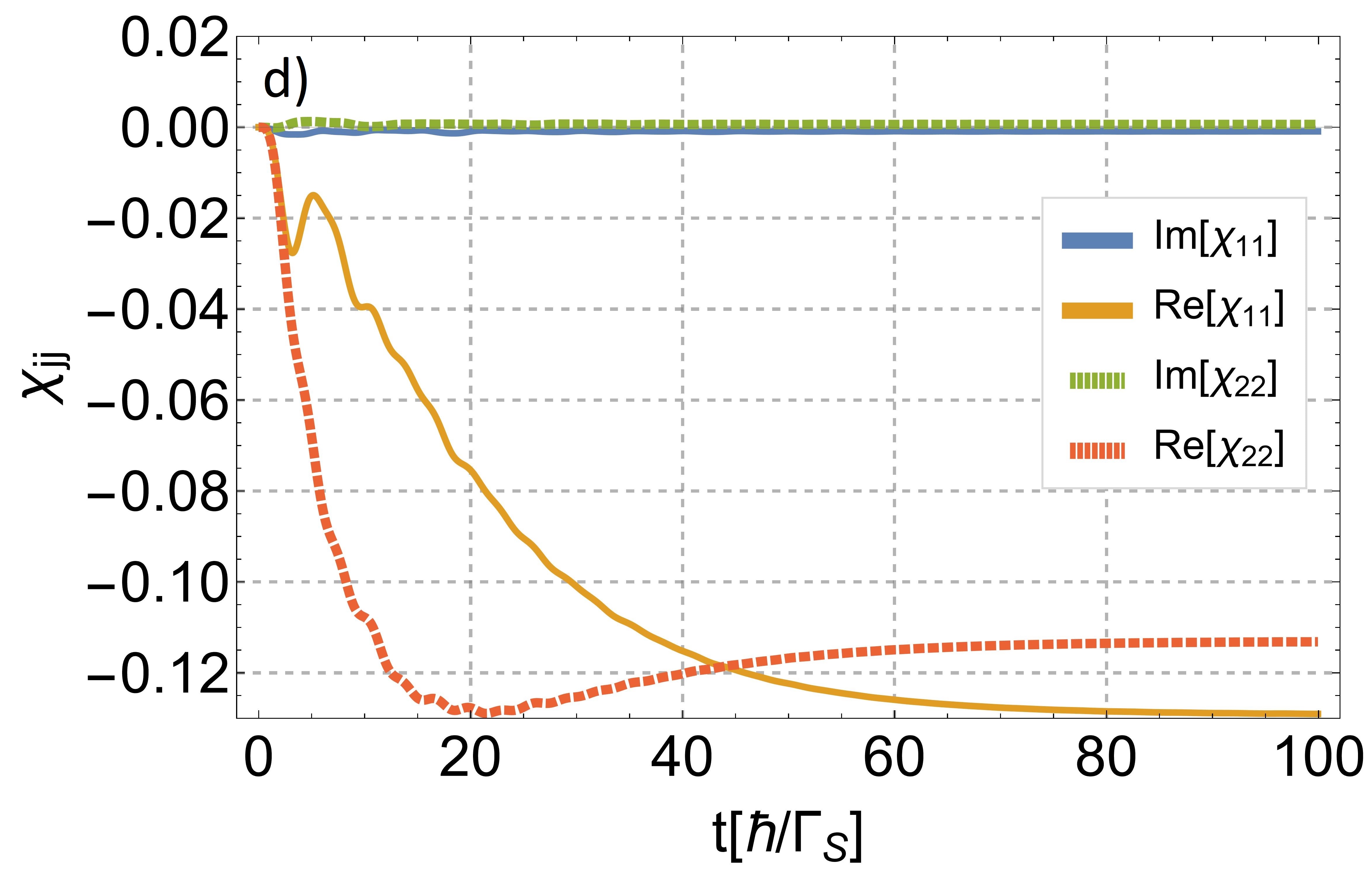}
\caption{The time-dependent electron occupancy of the quantum dots (panels a-c) and the pairing amplitudes $\chi_{jj}$ (panel d), obtained for $V_{12}=\Gamma_S$, $\varepsilon_{j\sigma}=0$, $\Gamma_N=0.2\Gamma_S$, 
assuming the initial conditions $n_{j \downarrow}=0$, $n_{j \uparrow}=1$. Panels b and c  show the evolution of $n_{j\sigma}(t)$ in the interval $t\in \left< 0, 5\right> \frac{\hbar}{\Gamma_S}$.}
\label{fig.t1}
\end{figure}

In Fig.~\ref{fig.t1} we show evolution of the time-dependent charge occupancies (see the upper and middle panels) and development of the on-dot electron pairing amplitudes obtained numerically for $V_{12}=\Gamma_{S}$, $\varepsilon_{1\sigma}=\varepsilon_{2\sigma}=0$. Gradual disappearance of the triplet configuration is possible via inflow/outflow of the normal metallic lead electrons to/from QD$_{2}$. This process is accompanied by gradual buildup of the on-dot pairing amplitudes over the time interval $\tau \sim \hbar/\Gamma_{N}$ (see the bottom panel). 

We observe, that charge redistribution reveals some tiny steps in the spin-dependent sectors. Their origin is strictly related with two-step charge transfer processes: first from N to QD$_{2}$, and next between the dots.
In the strongly coupled quantum dots, the spin-$\downarrow$ electron arriving from normal electrode quickly propagates further to QD$_1$, while spin-$\uparrow$ vacancy in QD$_2$ triggers a transfer of the spin-$\uparrow$ electron from QD$_1$ to QD$_2$.
Such electron transfer between the quantum dots occurs on a time scale much shorter than the tunneling process from $N$ to QD$_2$. In consequence, the occupancies of the dots exhibit rapid intertwined oscillations (Fig.~\ref{fig.t1}a-c) before stabilizing over the longer time scale ${\tau \sim \hbar/\Gamma_{N}}$, when the system  approaches the steady-state configuration $n_{j \sigma} \approx 1/2$.

To clarify the mechanism responsible for the intertwined oscillations it is useful to consider the steady limit ($t\rightarrow\infty$) solution for our setup. For analytical evaluations it is instructive to neglect the coupling to the normal metallic lead, $\Gamma_{N}=0$, when the single particle Green's function of both quantum dots can determined analytically. Fourier transform of the  QD$_1$ Green's function is given by 
\begin{eqnarray}
&&\langle\langle \hat{c}_{1\uparrow};\hat{c}_{1\uparrow}^{\dagger}\rangle\rangle
\label{eqn_16} \\&=&
\frac{\omega+\varepsilon_{1\downarrow}-\frac{(V_{12})^{2}}{\omega+\varepsilon_{2\downarrow}} } { \left(\omega-\varepsilon_{1\uparrow}-\frac{(V_{12})^{2}}{\omega-\varepsilon_{2\uparrow}}\right)\left(\omega+\varepsilon_{1\downarrow}-\frac{(V_{12})^{2}}{\omega+\varepsilon_{2\downarrow}}\right) - \left( \frac{\Gamma_{S}}{2}\right)^{2}} ,
\nonumber
\end{eqnarray}
implying four quasiparticle states [poles of Eqn.~(\ref{eqn_16})]. For the chosen quantum dot levels, $\varepsilon_{1\sigma}=0=\varepsilon_{2\sigma}$, these quasiparticles appear at energies
\begin{equation}
\omega = \frac{1}{2} \left( \pm \frac{\Gamma_S}{2} \pm \sqrt{\left(\frac{\Gamma_S}{2}\right)^{2}+4(V_{12})^{2}} \right)    .
\label{eqn17}
\end{equation}
The same molecular states, though with different spectral weights, are found for QD$_{2}$.

Specifically, for $V_{12}=\Gamma_S$ (Fig.~\ref{fig.t1}) the energies (\ref{eqn17}) are approximately equal to $\pm 1.28\Gamma_S$ and $\pm 0.78 \Gamma_{S}$. Gradual buildup of the molecular spectrum in the transient region (under nonequilibrium conditions) is accompanied by quantum oscillations between these emerging quasiparticle modes, as discussed by us in Ref.\ \cite{Taranko-2021}. For the chosen set of model parameters the fastest oscillatory behavior comes from transitions between the outer states of the frequency $\Omega = \frac{2\pi}{T} \approx 2*1.28$ with the period $T\approx 2.45$ [in units $\hbar/\Gamma_{S}$]. This periodicity of the intertwined oscillations is seen in Fig.~\ref{fig.t1}a-c.

\subsection{Weak inter-dot coupling}
\label{transient_weak}

    \begin{figure}[b!]
\includegraphics[width=0.95\linewidth]{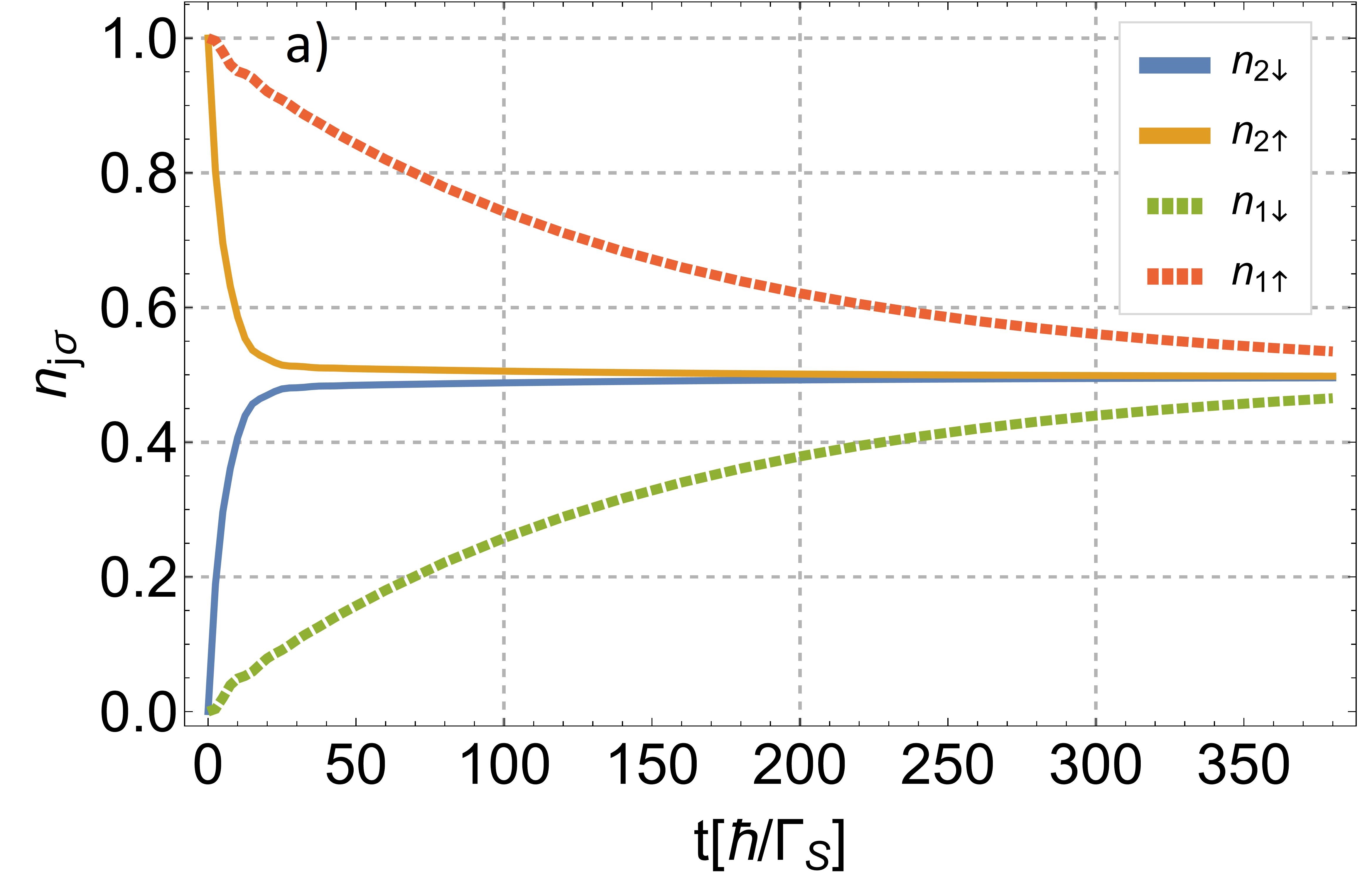}
\includegraphics[width=0.95\linewidth]{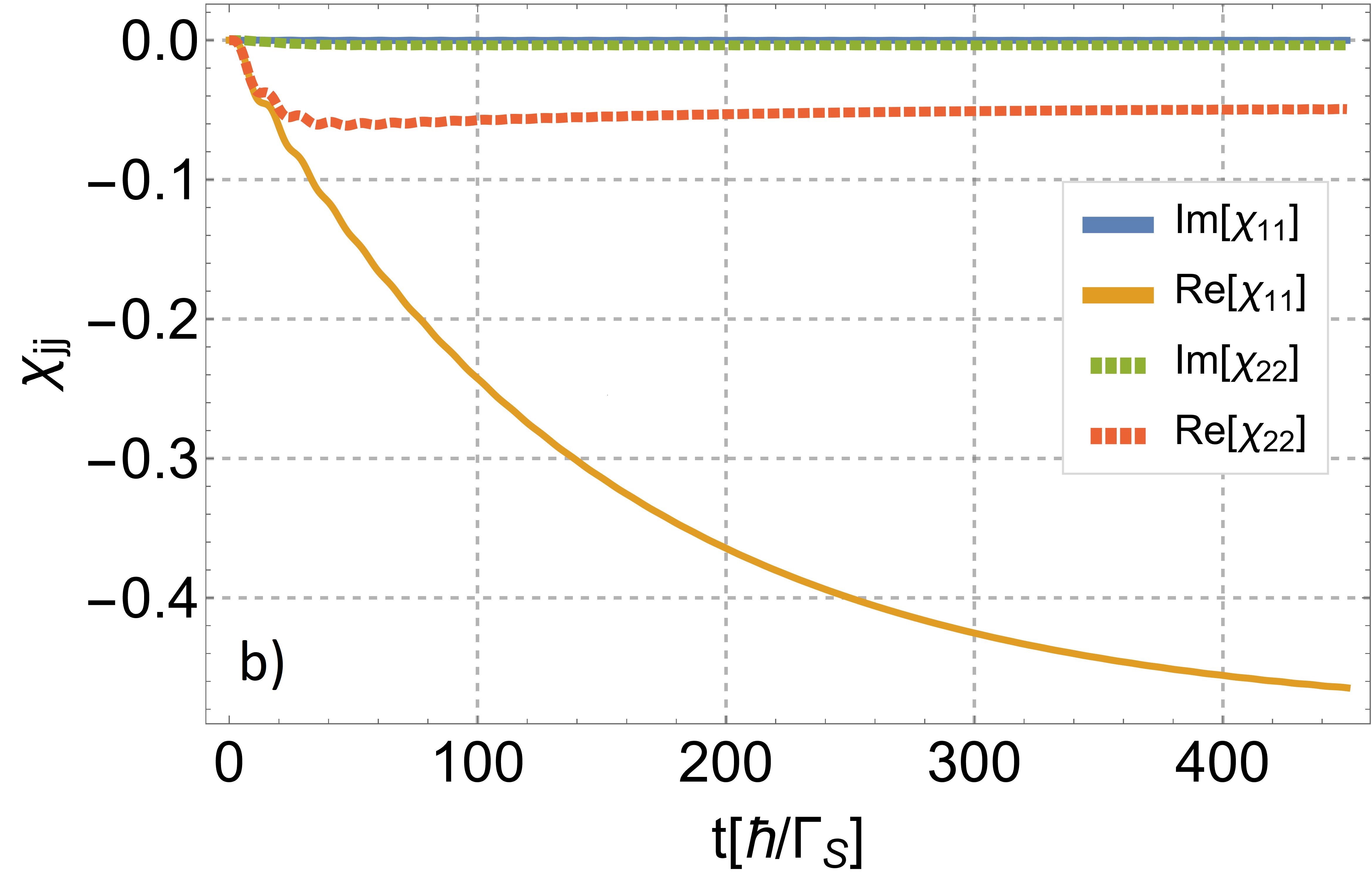}
\caption{The time-dependent electron occupancies of the quantum dots obtained in the weak coupling limit, $V_{12}=0.1\Gamma_S$, using the same model parameters as in Fig.~\ref{fig.t1}.
}
\label{fig.t01}
\end{figure}

\TD{Duration of the triplet configuration} looks differently when the inter-dot coupling is much weaker than the hybridization of QD$_2$ with the normal metallic lead, $V_{12}\ll\Gamma_N$. Charge transfer between the dots occurs then on a slower time scale than the tunneling processes between the normal metallic lead and QD$_2$. Under such conditions the rapid intertwined oscillations are missing.

Figure \ref{fig.t01} shows the time-dependent occupancies of both quantum dots obtained for $\varepsilon_{j\sigma}= 0$, assuming the initial triplet configuration (\ref{QD_occup_def}) and weak inter-dot coupling $|V_{12}|=0.1\Gamma_S$. Occupancy of QD$_2$ evolves to its steady limit value $n_{2\sigma}(t\rightarrow\infty)=1/2$ on the time scale $\tau \sim \frac{\hbar}{\Gamma_N}$, in analogy to the previously discussed (strong inter-dot coupling) case. In contrast, the time-dependent occupancy of QD$_1$ establishes over much longer time scale \TD{$\tau_{A}\sim \frac{\hbar}{V_{12}}$ because of slow charge transfer between the weakly-coupled quantum dots. The latter process, however, is crucial for developing the electron pairing on QD$_{1}$ which practically occurs after the time interval $\tau_{A}$ longer than $\tau$, characterizing the transient phenomena in QD$_{2}$. These distinct time scales are well evident in evolution of the occupancies (Fig.~\ref{fig.t01}a) and the on-dot pairing amplitudes (Fig.~\ref{fig.t01}b).}

We can also notice that the steady limit pairing amplitudes $|\chi_{ii}(t\rightarrow\infty)|$ are very different for individual quantum dots. The electron pairing slowly induced on QD$_1$ (which is directly coupled to superconductor) is  stronger in comparison to the quickly developed pairing of QD$_2$. This is a consequence of the weak inter-dot hybridization, $V_{12}\ll \Gamma_{S}$, which would affect the subgap Andreev transfer in the biased setup.

\section{Transient effects due to magnetic field and biasing}
\label{Zeeman.blockade}

We now consider another scenario in which the triplet \TD{configuration} could be enforced by applying the magnetic field \cite{Winkelmann_2020,Moehle_2021}, which would affect the in-gap bound states via spin degree of freedom \cite{Huang-2021,spin_filtered_Andreev-2023}. We assume that the external field $B$ is switched on at time instant $t_{0}$, imposing the Zeeman splitting of the energy levels
\begin{eqnarray}
\varepsilon_{j\uparrow}-\varepsilon_{j\downarrow}=
\left\{ \begin{array}{ll}
0 & \hspace{0.3cm} \mbox{\rm for } t\leq t_0 , \\
\mu_{B}B & \hspace{0.3cm} \mbox{\rm for } t> t_0 ,
\end{array} \right.
\end{eqnarray}
where $\mu_{B}$ is the Bohr magneton. In the unbiased (and infinitesimally biased) junction such quantum quench leads to polarization of the quantum dots. We study the post-quench evolution, investigating its signatures which could be observable in the subgap charge transport. 

For specific calculations we assume the quantum dots to be initially empty (at $t\leq 0$). Similarly to the previous section, we first analyze evolution of physical observables after forming S-QD$_1$-QD$_2$-N junction ($t>0$), and next  investigate influence of the magnetic field (for $t>t_0$). Finally, we check how robust is the triplet configuration against nonequilibrium conditions. For this purpose we assume the junction to be biased at time instant $t_1>t_0$. The  subgap charge current turns out to be extremely sensitive, both to the Zeeman field and to the source-drain voltage $V$. In what follows we discuss these effects for the strong and weak inter-dot coupling, respectively.

\begin{figure}
\includegraphics[width=0.92\linewidth]{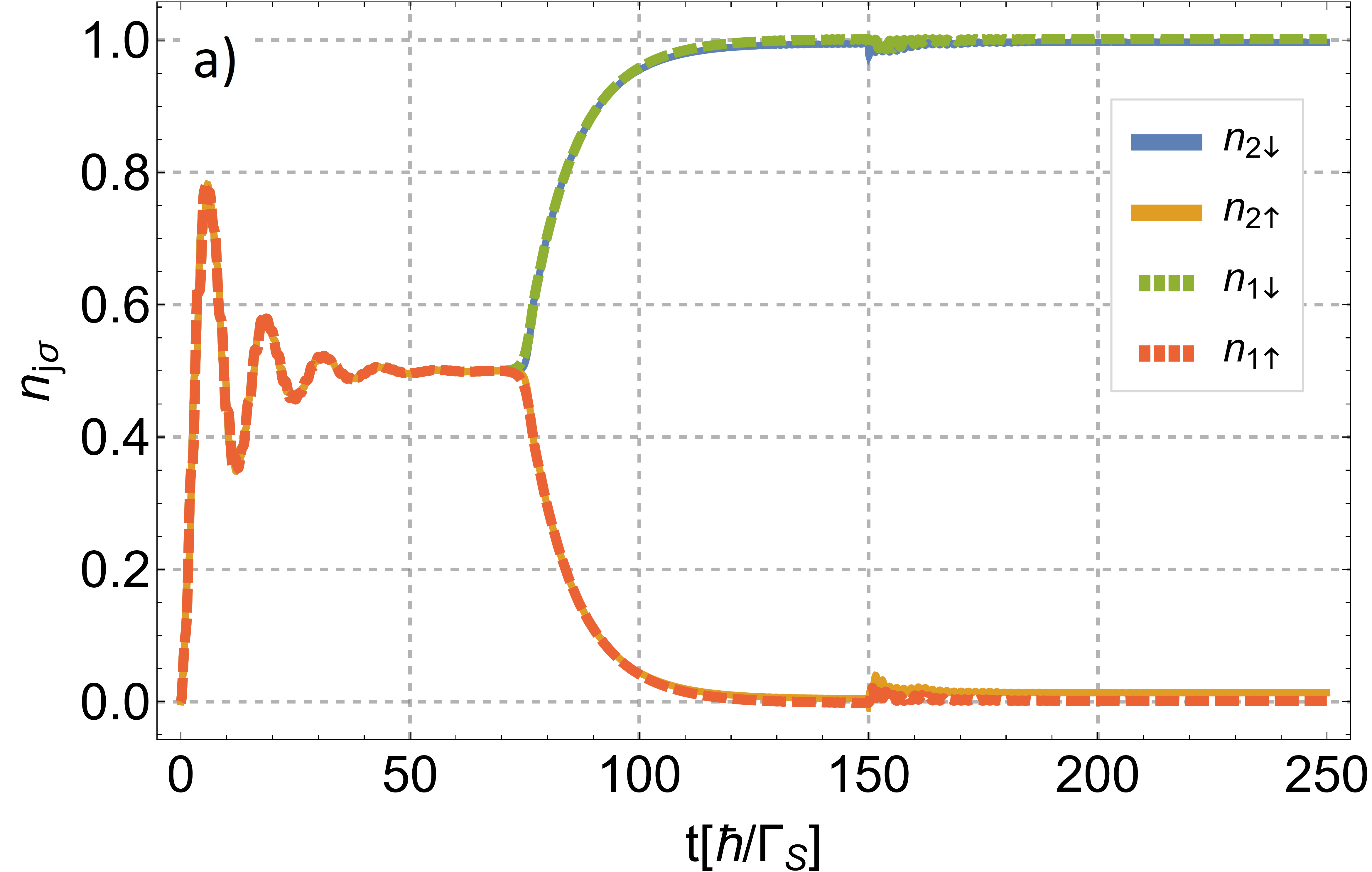}
\includegraphics[width=0.475\linewidth]{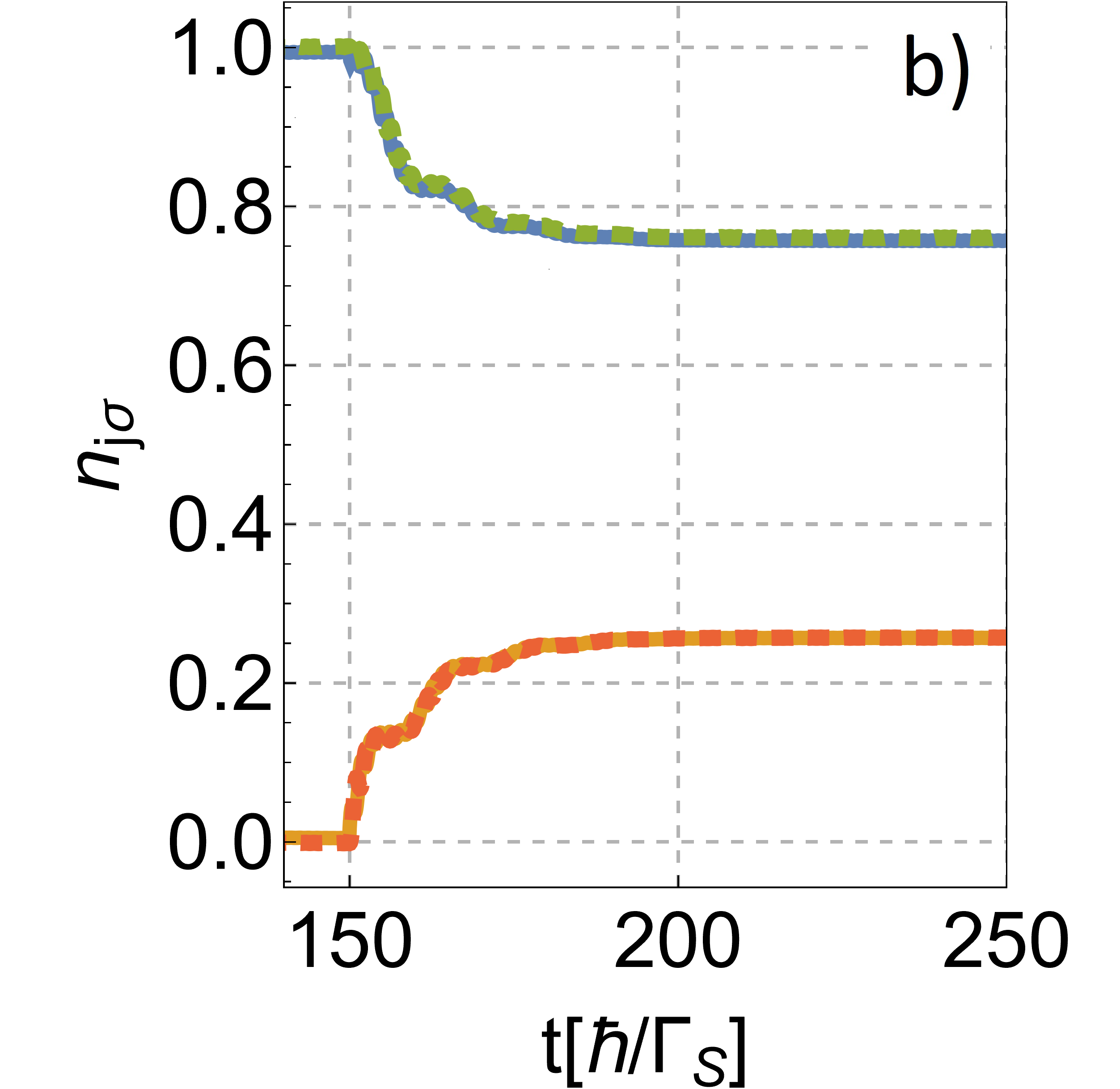}
\includegraphics[width=0.475\linewidth]{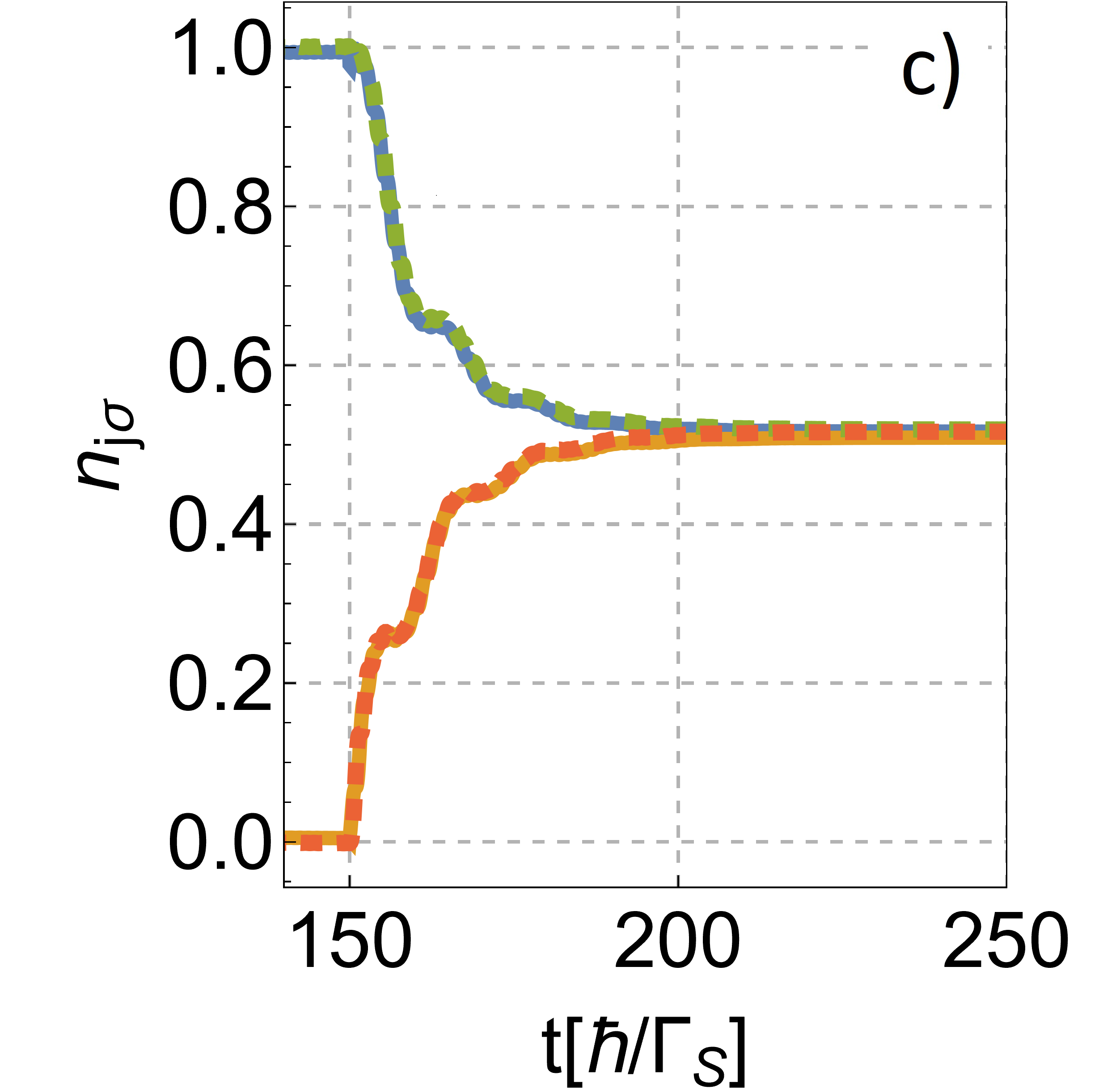}
\caption{The time-dependent occupancies of the quantum dots obtained for the empty quantum dots contacted at $t=0^+$ to external leads which, at $t_0=75\frac{\hbar}{\Gamma_s}$, are influenced by the magnetic field $B=10\Gamma_{S}/\mu_{B}$ forcing Zeeman split $\varepsilon_{j \uparrow}=5\Gamma_S$ and $\varepsilon_{j \downarrow}=-5\Gamma_S$. The junction is next biased at $t_1=150 \frac{\hbar}{\Gamma_s}$, using the source-drain voltage $V=2\Gamma_S$ (top panel), $V=5\Gamma_S$ (left-bottom panel) and $V=10\Gamma_S$ (right-bottom panel), respectively. Calculations have been performed for the strong inter-dot coupling,
$V_{12}=2\Gamma_{S}$, assuming $\Gamma_{N}=0.2\Gamma_{S}$. 
}
\label{fig.Zeeman_3pannel}
\end{figure}

\subsection{Strong inter-dot coupling}
\label{zeeman_strong}

\begin{figure}[b!]
\includegraphics[width=0.95\linewidth]{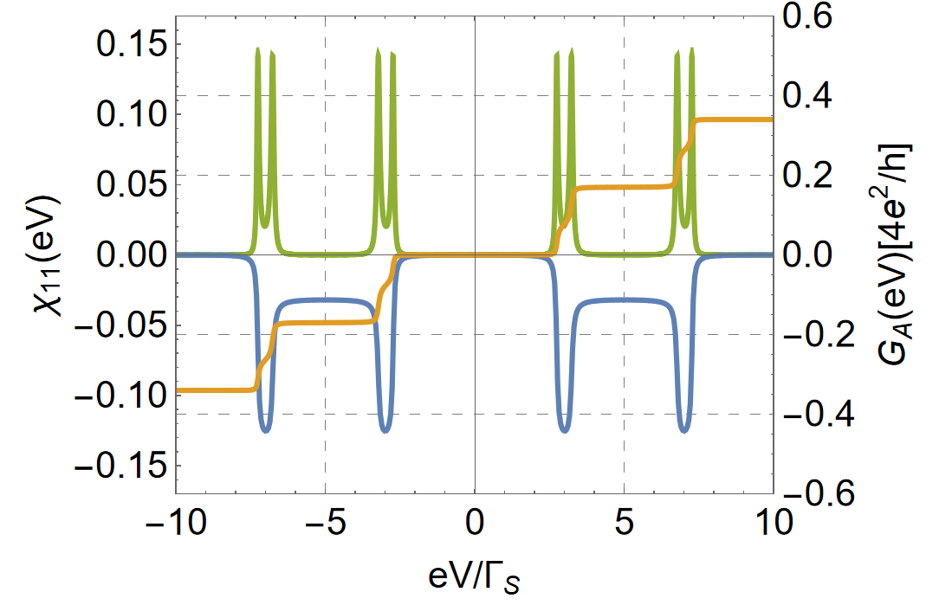}
\caption{The real/imaginary (blue/orange curve) parts of the electron pairing $\chi_{11}$ and the Andreev conductance (green curve) obtained in the steady limit ($t\rightarrow\infty$) for the model parameters $V_{12}=2\Gamma_S$, $\Gamma_{N}=0.2\Gamma_S$, 
and magnetic field $\mu_{B}B = 10\Gamma_S$, inducing Zeeman splitting $\varepsilon_{j \uparrow}=5\Gamma_S$, $\varepsilon_{j \downarrow}=-5\Gamma_S$. Results have been obtained from the expression analogous to Eqn.~(19) derived in Ref.\ \cite{Taranko-2021}.}
\label{fig.Zeeman_steady}
\end{figure}

Figure \ref{fig.Zeeman_3pannel} shows variation of the occupancy $n_{j\sigma}$ of the strongly coupled quantum dots obtained for $\varepsilon_{j\sigma} = 0$. They approach the half-filling $n_{j\sigma} \approx 1/2$ practically already at $t\rightarrow 50$. After applying the magnetic field, at $t_0 = 75$, the number of $\sigma = \downarrow$ ($\uparrow$) electrons of both dots exponentially increases (decreases) with time. After about $\tau\sim\hbar/\Gamma_{N}$, the triplet configuration is firmly established, $n_{j\uparrow}(t\geq \tau) \simeq 0$ and $n_{j\downarrow}(t\geq \tau) \simeq 1$. Its appearance is accompanied by suppression of the on-dot pairing amplitudes (see Fig.\ \ref{fig.Apendix_chi11} in Appendix \ref{app_Pairing_chi11}).

The detrimental influence of magnetic field on the electron pairing can $\chi_{jj}(t)$ be partly or entirely compensated by a voltage $V$ applied across the junction. To demonstrate this we assume that our setup is biased at $t_1=150\frac{\hbar}{\Gamma_S}$. Fig.\ \ref{fig.Zeeman_3pannel}a shows the results obtained for weak voltage, $V = 2\Gamma_S$, when the triplet configuration is almost unchanged. In contrast, in the strongly biased case (Fig.\ \ref{fig.Zeeman_3pannel}b,c) spin-dependent populations of the quantum dots become substantially rearranged. In consequence, sufficiently large voltage enforces the subgap current.

To clarify why the infinitesimally small voltage does not induce any noticeable charge current it is worth to analyze the steady limit pairing $\chi_{11}(t \rightarrow \infty)$ and the differential conductance (Fig.\ \ref{fig.Zeeman_steady}) where peaks of $G_{A}(V)$ indicate positions of the bound states. The subgap current through weakly biased junction is missing if none of the in-gap bound states enters the transport window, $\mu_{N}-\mu_{S}=V$. We encounter here such effect, resembling the blockade discussed discussed in Refs.\ \cite{Pekker-2021,Zhang-2022}. In our case, however, the blockage is driven by the magnetic field instead of the correlations.

\begin{figure}
\includegraphics[width=0.95\linewidth]{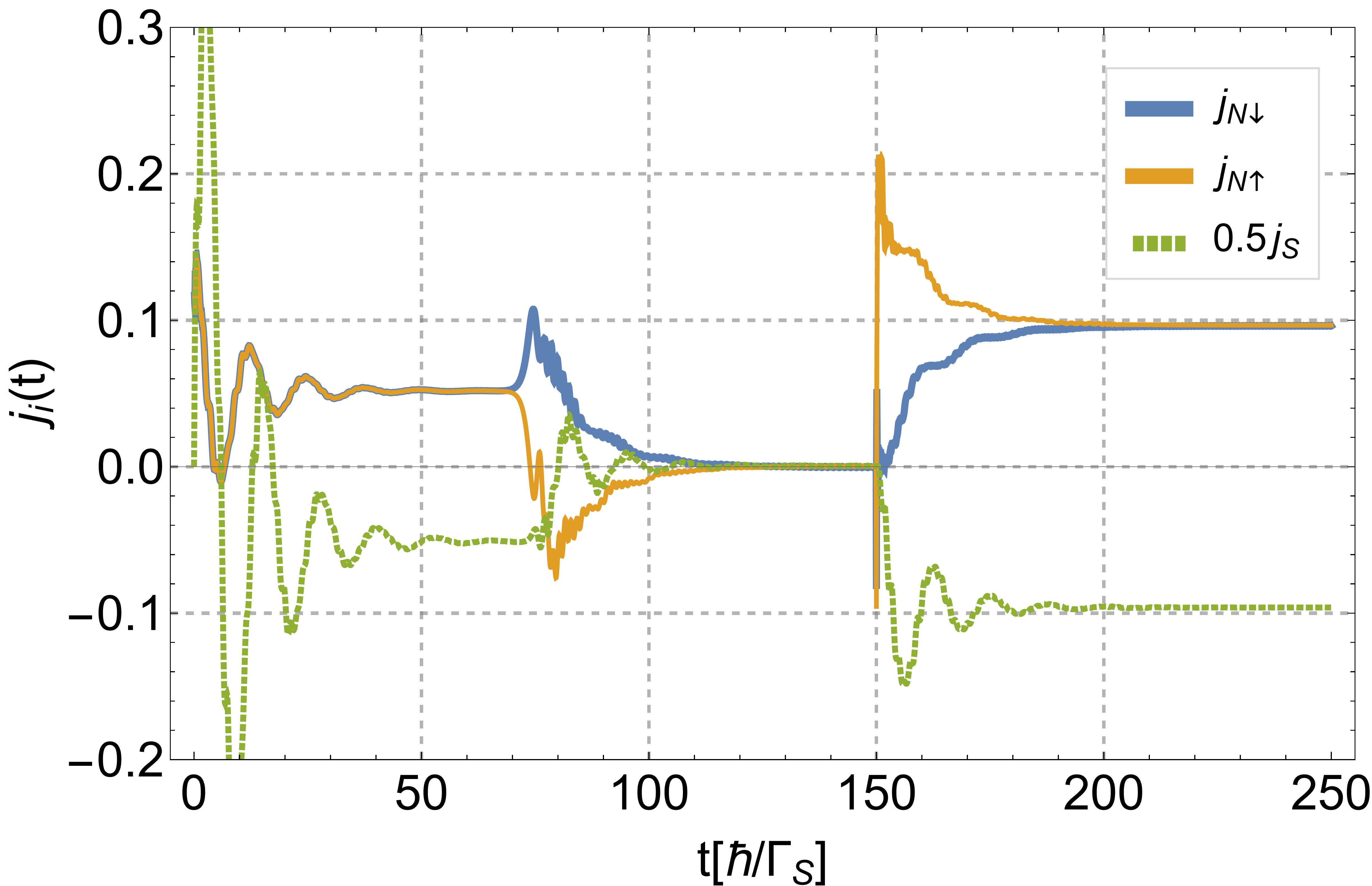}
\caption{The time-dependent charge currents flowing from the normal lead, $j_{N\sigma}$, and from superconducting electrode, $j_{S}$, through S-QD$_1$-QD$_2$-N junction weakly biased at $t=0^+$ by the initial voltage $V_0=2\Gamma_S$. At $t_0=75\frac{\hbar}{\Gamma_S}$ the magnetic field is switched on, imposing the triplet configuration and suppressing the charge currents. Next, at $t_1=150\frac{\hbar}{\Gamma_S}$, we increase the bias voltage to $V=10\Gamma_S$, enabling the subgap charge transport.}
\label{fig.Zeeman_current}
\end{figure}

Transfer of the spin-$\uparrow$ electrons from the normal metallic electrode to QD$_2$ is eventually possible when the voltage is sufficiently large (Fig.\ \ref{fig.Zeeman_current}). 
We evaluated the time-interval needed for disappearance of the subgap current blockade (due to the strong bias) by fitting our data to an exponential behavior, following the procedure used previously in Ref.\ \cite{Baranski_etal-2021}. For the chosen set of parameters we obtained that such characteristic time is $\tau \approx 10 \frac{\hbar}{\Gamma_S}$. For the hybridization values used in experiments \cite{Wernsdorfer-2012}, $\Gamma_{S} \sim 200 \,\mu\text{eV}$, this would yield $\tau \sim$ 34 ps.

\subsection{Weak inter-dot coupling}

\begin{figure}[b!]
\includegraphics[width=0.95\linewidth]{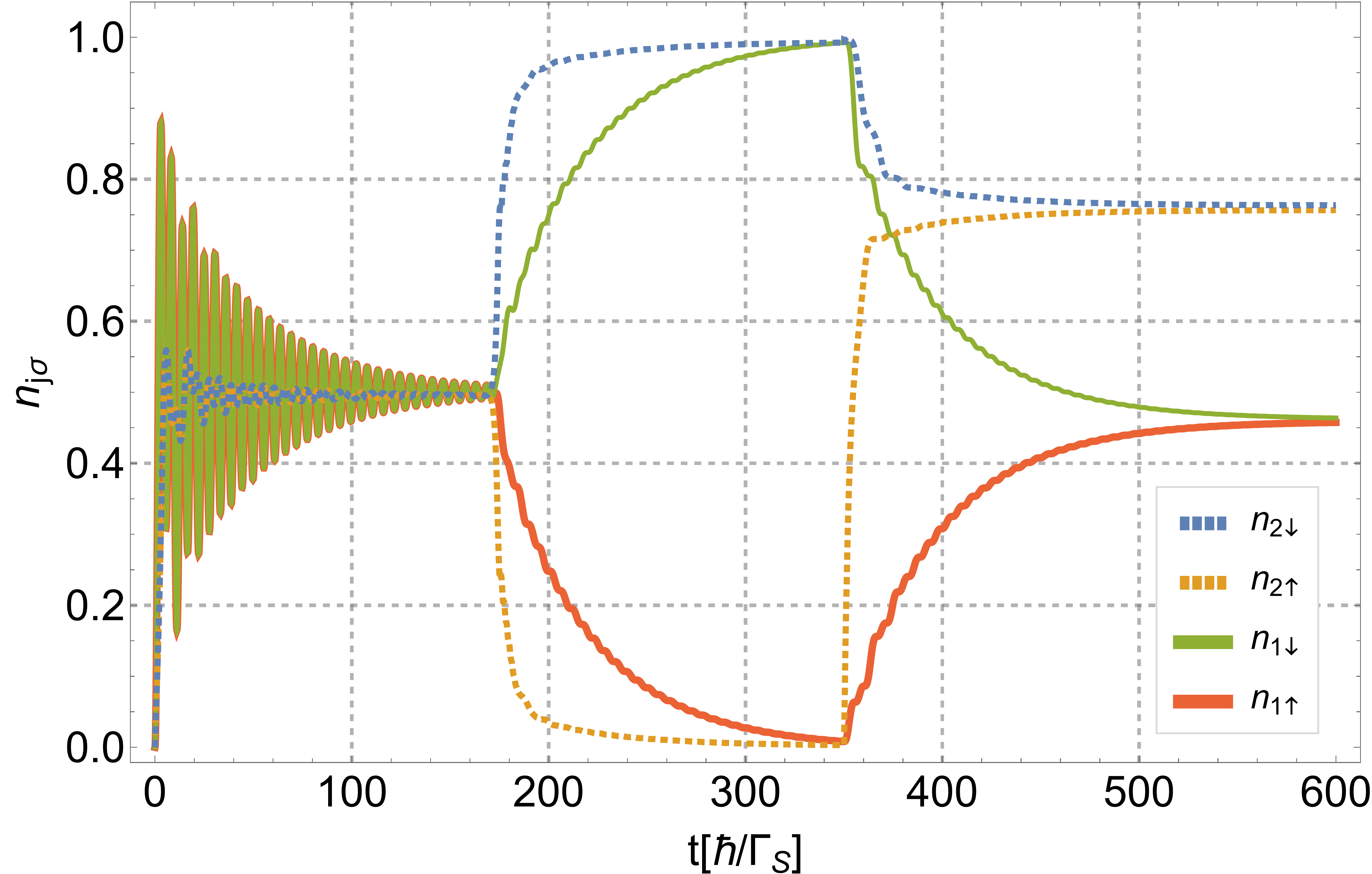}
\caption{Time-dependent occupancy obtained for the initially empty quantum dots which at $t=0^{+}$ are weakly coupled $V_{12}=0.2\Gamma_S$  and contacted to the external leads. Later on, at $t_0=180$, the Zeeman field is imposed inducing the triplet configuration. Finally, at $t_1=350$, the junction is biased by source-drain voltage $V=10\Gamma_S$.}
\label{fig.nt02}
\end{figure}

\begin{figure}[b!]
\includegraphics[width=0.95\linewidth]{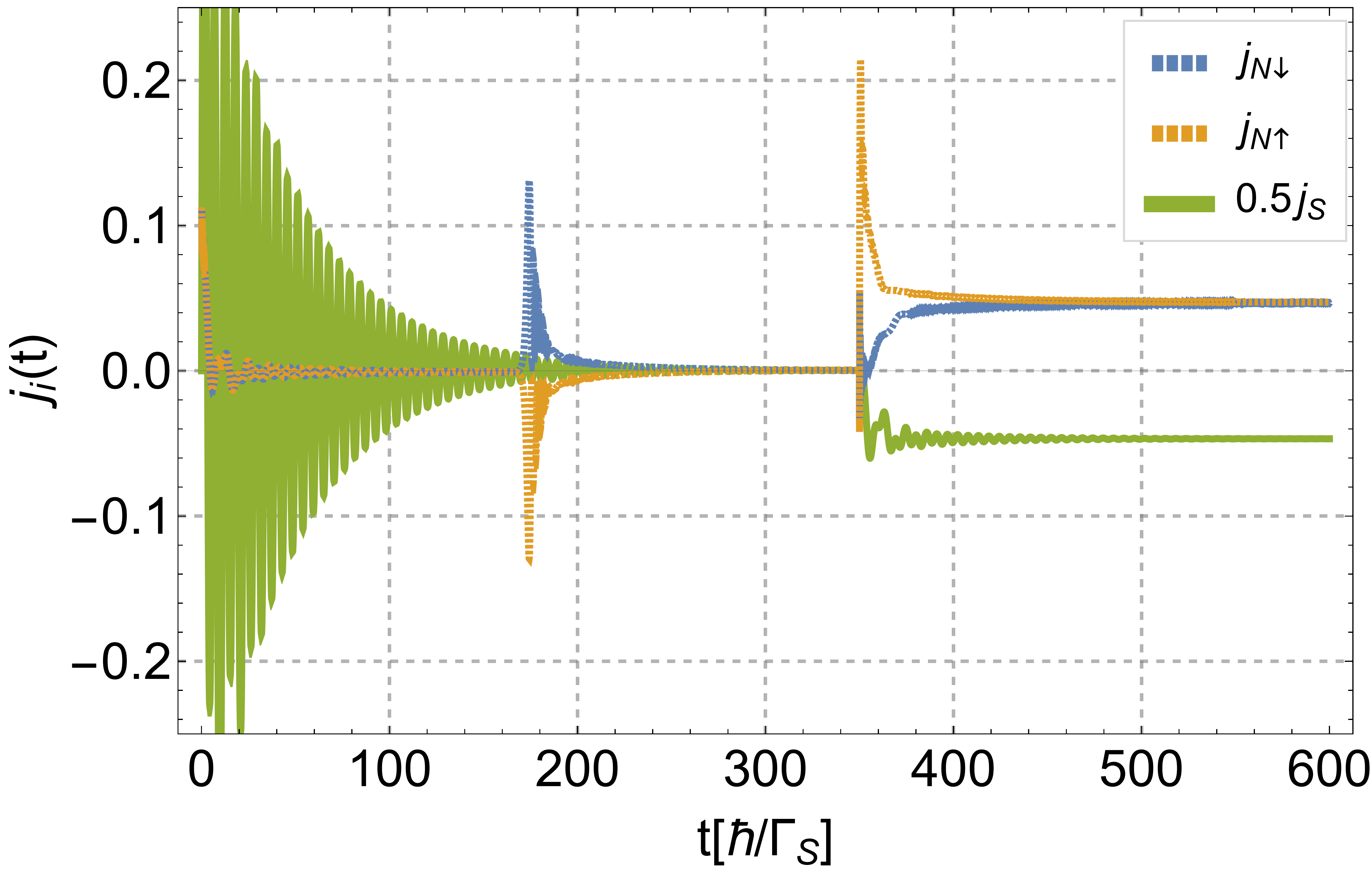}
\caption{Time-dependent currents obtained for the weakly
coupled quantum dots for the situation presented in Fig.~\ref{fig.nt02}.}
\label{fig.weak_current}
\end{figure}

In the case of weak inter-dot coupling the time-dependent phenomena driven by the Zeeman field and/or application of the source-drain voltage reveal two distinct time scales. One of them, $\tau \sim \hbar/\Gamma_N$,  specifies the time-interval for charge inflow/outflow between QD$_{2}$ and the metallic lead. The other time-interval, $\tau_A\sim \hbar/|V_{12}|$, characterizes much slower process for transmitting the charge between weakly coupled dots. The latter one controls how long it takes for reconstructing the in-gap bound states in response to the magnetic field, affecting the subgap current across the junction.

These different time scales can be well recognized in Fig.\
\ref{fig.nt02}, which shows variation of the electron occupancy $n_{j\sigma}(t)$ of the weakly coupled quantum dots, $V_{12}=0.2\Gamma_S$. In analogy to the transient effects discussed in Sec.~\ref{transient_weak}, we notice that the quantum dot QD$_2$ exchanges electrons with the metallic lead over the time interval $\tau$. Spin-dependent occupancy of QD$_1$  needs much more time, $\tau_A > \tau$, to achieve the new configuration imposed by the magnetic field. In the absence of any bias voltage, the quantum dots evolve into the triplet configuration, when the subgap charge transfer is suppressed.
This situation can be changed by applying the strong source-drain voltage. Fig.~\ref{fig.weak_current} shows that recovery of the subgap currents would occur within the time interval $\tau_A$.

\section{Summary}

We studied the transient effects related to the triplet configuration appearing in two quantum dots coupled in series between the normal metallic electrode and superconductor. We found that, the on-dot electron pairing is suppressed whenever both quantum dots are singly occupied by the same-spin electrons. In consequence, such configuration blocks the Andreev charge transport through this weakly biased system.

We analyzed in detail the characteristic time scales of the triplet state duration in double quantum dots right after connecting them to the external leads. From analytical expressions for the physical observables (such as spin-resolved occupancies, pairing amplitudes, and charge currents between constituents of this setup), we evaluated how much time it takes to eliminate the subgap blockade via  electron inflow/outflow from/to the metallic lead. Furthermore, we showed that such processes qualitatively depend on the inter-dot coupling, $V_{12}$. In the strongly coupled dots the charge redistribution occurs practically simultaneously in both quantum dots, leading to formation of their molecular bound states. Under such circumstances, persistence of the triplet configuration is merely controlled by the hybridization of QD$_{2}$ with the metallic electrode, $\tau \sim \hbar/\Gamma_N$. In contrast, in the weakly coupled quantum dots the triplet blockade evolves differently. Electron population of the quantum dot neighboring to the metallic lead establishes after the time interval $\tau\sim\hbar/\Gamma_N$, whereas the other dot (next to superconductor) evolves to its steady-state occupancy over much longer time scale, roughly given by $\tau_{A}\sim\hbar/|V_{12}|$. The latter process plays decisive role for lifting the triplet configuration enabling the on-dot pairing to be induced.

We considered also the other (empirically feasible) scenario of
imposing the triplet configuration by suddenly applying the magnetic field. We investigated the post-quench observables,
determining the spin-resolved occupancies, pairing amplitudes and transient currents flowing via all segments of the setup. We obtained one characteristic time scale, $\tau\sim\hbar/\Gamma_N$, after which the triplet configuration establishes in the strongly hybridized dimer, whereas for the case of weakly coupled quantum dots two different time intervals, $\tau_A >\tau$, characterize two-stage emergence of such configuration.

Furthermore, we demonstrated that the triplet configuration can be partially or completely eliminated by biasing the junction.
In the linear response limit (i.e. for infinitesimally small voltage) this configuration is practically unaffected. Triplet state (induced in our case by magnetic field) is thus blocking any subgap charge transport, in analogy to the Andreev blockade discussed in Refs.\ \cite{Pekker-2021,Zhang-2022}. For the 
stronger source-drain voltage, the transport window eventually captures a fraction or entire spectrum of the in-gap bound states, enforcing the Andreev tunneling. Thus under the strong nonequilibrium conditions the transient as well as steady subgap current can pass through the junction.

Dynamics of the triplet configuration studied in this paper might be relevant to variety of superconducting nanohybrid structures, in particular, those where the superconducting qubits (formed from the in-gap states) are manipulated in non-adiabatic processes. The characteristic time scales predicted by our study could be encountered under the nonequilibrium situations, either due to varying the external magnetic fields, gate potentials, or source-drain biasing.  

\begin{acknowledgements}
This research project has been supported by the National
Science Centre (Poland) through the grant No.~2022/04/Y/ST3/00061.
\end{acknowledgements}

\appendix


\section{Laplace transforms}
\label{app_Laplace}

For computing the expectation values of the charge occupancy and the local electron pairing in the quantum dots we used the following Laplace transforms (valid for $U=0$)

\onecolumngrid
\begin{eqnarray}
\hat{c}_{1\uparrow}(s)&=&
\left. \hat c_{1 \uparrow}(0) \frac{(s+i\varepsilon_{2 \uparrow}+g)A_1(s)}{A_3(s)}\right.  -i\Delta \hat c_{1\downarrow}^{\dagger}(0)\frac{(s+i\varepsilon_{2 \uparrow}+g)(s-i\varepsilon_{2 \downarrow}+g)}{A_{3}(s)}
-iV_{12} \hat c_{2 \uparrow}(0) \frac{A_1(s)}{A_3(s)} \nonumber \\
&+&   \Delta V_{12} \hat c_{2 \downarrow}^{\dagger}(0) \frac{s+i\varepsilon_{2 \uparrow}+g}{A_3(s)} +i\Delta V_{12} \sum_{\bf{k}} \frac{V_{N{\bf k}} \hat c_{N{\bf k} \downarrow}^{\dagger}(0) }{s-i\xi_{N{\bf k}}}\frac{(s+i\varepsilon_{2 \uparrow}+g)}{A_3(s)} -V_{12} \sum_{\bf k} \frac{V_{N{\bf k}} \hat c_{N{\bf k} \uparrow}(0)}{s+\xi_{N{\bf k}}}\frac{A_1(s)}{A_3(s)},
\label{c1ups}
\end{eqnarray}
\begin{eqnarray}
\hat c_{1 \downarrow}(s) &=& i \Delta \hat c_{1 \uparrow}^{\dagger}(0) \frac{(s+i\varepsilon_{2 \downarrow}+g)(s-i\varepsilon_{2 \uparrow}+g)}{A_3^*(s)} + \hat c_{1 \downarrow}(0)\frac{(s+i\varepsilon_{2 \downarrow}+g) A_2^*(s)}{A_3^*(s)}  -i V_{12} \hat c_{2 \downarrow}(0)\frac{A_2^*(s)}{A_3^*(s)} \nonumber \\
&-& \Delta V_{12} \hat c_{2 \uparrow}^{\dagger} \frac{s+i\varepsilon_{2 \downarrow}+g}{A_3^*(s)} -V_{12} \sum_{\bf k} \frac{V_{N{\bf k}}\hat c_{N{\bf k} \downarrow}(0) }{s+i\xi_{N{\bf k}}}\frac{A_{2}^*(s)}{A_3^*(s)} -i \Delta V_{12} \sum_{\bf k} \frac{V_{N{\bf k}} \hat c_{N{\bf k} \uparrow}}{(s-i\xi_{N{\bf k}})}\frac{(s+i\varepsilon_{2 \downarrow}+g)}{A_3^*(s)},
\label{c1dos}
\end{eqnarray}
\begin{eqnarray}
\hat{c}_{2\uparrow}(s)&=& \hat{c}_{2\uparrow}(0)\frac{A_3(s)-V_{12}^2A_{1}(s)}
{(s+i\varepsilon_{2\uparrow}+g)A_3(s)} -i \hat{c}_{2\downarrow}^{\dagger}(0) 
\frac{\Delta V_{12}^{2}}{A_{3}(s)} -\hat{c}_{1\downarrow}^{\dagger}(0) \frac{V_{12}\Delta (s-i\varepsilon_{2\downarrow}+g)}{A_{3}(s)} \nonumber \\ 
&-&i\hat{c}_{1\uparrow}(0) \frac{V_{12}A_{1}(s)}{A_{3}(s)} + \Delta V_{12}^{2} \sum_{\bf k} V_{N{\bf k}} 
\frac{\hat{c}_{N{\bf k}\downarrow}^{\dagger}(0)}{(s-i\xi_{N{\bf k}})A_3(s)} - i\sum_{\bf k} V_{N{\bf k}} \frac{\hat{c}_{N{\bf k}\uparrow}(0)[A_3(s)-V_{12}^2 A_{1}(s)]}{(s+i\varepsilon_{N{\bf k}})(s+i\varepsilon_{2\uparrow}+g)A_3(s)},
\label{c2ups}
\end{eqnarray}
\begin{eqnarray}
    \hat{c}_{2\downarrow}(s)&=& \frac{\hat{c}_{2\downarrow}(0) [A_3^*(s)-V_{12}^2 A_2^*(s)]}
{(s+i\varepsilon_{2\downarrow}+g)A^*_{3}(s)} + \hat{c}_{2\uparrow}^{\dagger}(0)
 \frac{i\Delta V_{12}^{2}}{A_{3}^{*}(s)} -\hat{c}_{1\downarrow}(0) \frac{iV_{12}A_{2}^{*}(s)}{A_{3}^{*}(s)} 
+\hat{c}_{1\uparrow}^{\dagger}(0) \frac{\Delta V_{12}(s-i\varepsilon_{2\uparrow}+g))}{A_{3}^{*}(s)} \nonumber \\
&-& i\sum_{\bf k} V_{N{\bf k}}  \frac{\hat{c}_{N{\bf k}\downarrow}(0)[A_{3}^*(s)-V_{12}^2A_{2}^*(s)]}
{(s+i\xi_{N{\bf k}})(s+i\varepsilon_{2\downarrow}+g)A_3^*(s)} - \Delta V_{12}^{2}\sum_{\bf k} V_{N{\bf k}} \frac{\hat{c}_{N{\bf k}\uparrow}^{\dagger}(0)}{(s-i\xi_{N{\bf k}})A_{3}^{*}(s)} .
\label{c2dos}
\end{eqnarray}
where $g=\Gamma_{N}/2$, $\Delta=\Gamma_{S}/2$ and $A_{i}(s)$ are defined in Eqs.~(\ref{Aa_1},\ref{Aa_2},\ref{Aa_3}).

\section{Time-dependent observables}
\label{observables}

Using the Laplace transforms presented in Appendix \ref{app_Laplace} we can determine the spin-resolved electron occupancies
$n_{j\sigma}(t) = \left< {\cal{L}}^{-1} \left\{\hat{c}_{j\sigma}^{\dagger}(s)
\right\}(t){\cal{L}}^{-1} \left\{\hat{c}_{j\sigma}(s)\right\} (t)\right>$
of both quantum dots. Here we present their expressions for $n_{1 \downarrow}$, $n_{2 \sigma}$
\begin{eqnarray}
n_{1 \downarrow}(t)&=&n_{1 \downarrow}(0)\left|{\cal{L}}^{-1}\left\{\frac{(s-i\varepsilon_{2 \downarrow}+g)A_2(s)}{A_3(s)} \right\}(t)\right|^2 
    +n_{2 \downarrow}(0) V_{12}^2\left|{\cal{L}}^{-1}\left\{\frac{A_2(s)}{A_3(s)} \right\}(t)\right|^2 \nonumber \\
    && + [1-n_{1 \uparrow}(0)] \Delta^2 \left|{\cal{L}}^{-1}\left\{\frac{(s-i\varepsilon_{2 \downarrow}+g)(s+i\varepsilon_{2 \uparrow}+g)}{A_{3}(s)} \right\}(t)\right|^2 \nonumber \\
    &&+[1-n_{2 \uparrow}(0)]\Delta^2V_{12}^2\left|{\cal{L}}^{-1}\left\{\frac{s-i\varepsilon_{2 \downarrow}+g}{A_3(s)} \right\}(t)\right|^2 \nonumber \\
    &&+\frac{\Gamma_N}{2\pi}V_{12}^2 \int_{-\infty}^{+\infty}d\varepsilon f_{N}(\varepsilon) \left|{\cal{L}}^{-1}\left\{ \frac{A_2(s)}{(s- i\varepsilon)A_{3}(s)} \right\}(t)\right|^2 \nonumber \\
    &&+ \frac{\Gamma_N}{2\pi}\Delta^2 V_{12}^2 \int_{-\infty}^{+\infty} d\varepsilon [1-f_{N}(\varepsilon)]\left|{\cal{L}}^{-1}\left\{ \frac{(s-i\varepsilon_{2 \downarrow}+g)}{(s+i\varepsilon)A_3(s)}\right\}(t)\right|^2,
    \label{n1downt}
\\
n_{2\uparrow}(t) &=& 
 \left( 1- n_{1\downarrow}(0) \right) \Delta^{2}V_{12}^{2} \left|
{\cal{L}}^{-1} \left\{ \frac{(s-i\varepsilon_{2\downarrow}+g)}
{A_{3}(s)} \right\} (t) \right|^{2}
+ \left( 1- n_{2\downarrow}(0) \right) \Delta^{2} V_{12}^{4} \left| {\cal{L}}^{-1} 
\left\{ \frac{1} {A_{3}(s)} \right\} (t) \right|^{2} \nonumber \\ 
&+& n_{1\uparrow}(0) V_{12}^{2} \left| {\cal{L}}^{-1} 
\left\{ \frac{A_{1}(s)}{A_{3}(s)} \right\} (t) \right|^{2}
+n_{2\uparrow}(0)  \left| {\cal{L}}^{-1} 
\left\{ \frac{1}{s+i\varepsilon_{2\uparrow}+g} 
-\frac{V_{12}^{2}A_{1}(s)}{A_{3}(s)(s+i\varepsilon_{2\uparrow}+g)}
\right\} (t) \right|^{2}
\nonumber \\ & + & \Gamma_{N} \frac{\Delta^{2}V_{12}^{4}}{2\pi}
\int_{-\infty}^{+\infty} d\varepsilon \left[ 1-f_{N}(\varepsilon)\right] \left|
{\cal{L}}^{-1} \left\{ \frac{1}
{(s-i\varepsilon)A_{3}(s)}\right\} (t) \right|^{2}
 \nonumber \\
&+& \frac{\Gamma_{N}}{2\pi} \int_{-\infty}^{+\infty} d\varepsilon f_{N}(\varepsilon) \left|
{\cal{L}}^{-1} \left\{ \frac{V_{12}^{2}A_{1}(s)}
{(s+i\varepsilon)(s+i\varepsilon_{2\uparrow}+g)A_{3}(s)}
-\frac{1}{(s+i\varepsilon)(s+i\varepsilon_{2\uparrow}+g)}
\right\} (t) \right|^{2} , 
\label{n_2up}
\end{eqnarray}  
\begin{eqnarray}
n_{2 \downarrow}(t)&=&n_{1 \downarrow}(0)V_{12}^2 \left|{\cal{L}}^{-1}\left\{ \frac{A_2(s)}{A_3(s)} \right\}(t)\right|^2 \nonumber +n_{2 \downarrow}(0) \left|{\cal{L}}^{-1}\left\{\frac{A_3(s)-V_{12}^2A_2(s)}{(s-i\varepsilon_{2 \downarrow}+g)A_3(s)} \right\}(t)\right|^2 \nonumber \\
&&+[1-n_{1 \uparrow}(0)]V_{12}^2\Delta^2\left|{\cal{L}}^{-1}\left\{\frac{s+i\varepsilon_{2 \uparrow}+g}{A_3(s)} \right\}(t)\right|^2+
[1-n_{2\uparrow}(0)] V_{12}^4\Delta^2\left|{\cal{L}}^{-1}\left\{\frac{1}{A_3(s)} \right\}(t)\right|^2\nonumber \\
&&+ \frac{\Gamma_N}{2\pi} \int_{-\infty}^{+\infty} d\varepsilon f_{N}(\varepsilon)\left|{\cal{L}}^{-1}\left\{\frac{A_3-V_{12}^2A_2(s)}{(s-i\varepsilon_{2 \downarrow}+g)(s-i\varepsilon)A_3(s)} \right\}(t)\right|^2 \nonumber \\
&&+\frac{\Gamma_N}{2\pi}\Delta^2V_{12}^4 \int_{-\infty}^{+\infty} d\varepsilon[1-f_{N}(\varepsilon)] \left|{\cal{L}}^{-1}\left\{ \frac{1}{(s+i\varepsilon)A_3(s)} \right\}(t)\right|^2,
\end{eqnarray}
where $f_{N}(\varepsilon)=\left[1+\mbox{\rm exp}\left( \frac{\varepsilon-\mu_{N}}{k_{B}T}\right)\right]^{-1}$.
Similar procedure allows us to determine the time-dependent pairing functions
\begin{eqnarray}
    \langle c_{1 \downarrow}(t) c_{1 \uparrow}(t)\rangle&=&
    n_{1 \uparrow}(0) i\Delta {\cal{L}}^{-1}\left\{ \frac{(s+ i\varepsilon_{2 \downarrow}+g)(s-i\varepsilon_{2 \uparrow}+g)}{A_3^*(s)} \right\}(t){\cal{L}}^{-1}\left\{ \frac{A_1(s)(s+i\varepsilon_{2 \uparrow}+g)}{A_3(s)}\right\} (t) \nonumber \\
    &&+n_{2 \uparrow}(0) i V_{12}^2 \Delta {\cal{L}}^{-1}\left\{ \frac{s+i\varepsilon_{2 \downarrow}+g}{A_3^*(s)}\right\}(t){\cal{L}}^{-1}\left\{ \frac{A_1(s)}{A_3(s)} \right\}(t) \nonumber \\
    &&-[1-n_{1 \downarrow}(0)]i\Delta {\cal{L}}^{-1}\left\{\frac{(s+i\varepsilon_{2 \downarrow}+g)A_2^*(s)}{A_3^*(s)} \right\}(t){\cal{L}}^{-1}\left\{ \frac{(s+i\varepsilon_{2 \uparrow} +g)(s-i\varepsilon_{2 \downarrow}+g)}{A_3(s)}\right\}(t)
    \nonumber \\
    && -i \Delta V_{12}^2[1-n_{2 \downarrow}(0)]{\cal{L}}^{-1}\left\{ \frac{A_2^*(s)}{A^*_{3}(s)} \right\}(t){\cal{L}}^{-1}\left\{ \frac{s+i\varepsilon_{2 \uparrow}+g}{A_{3}(s)} \right\}(t) \nonumber \\
   &&+i\Delta V_{12}^2 \frac{\Gamma_N}{2 \pi} \int_{-\infty}^{+\infty} d\varepsilon f_N(\varepsilon) {\cal{L}}^{-1}\left\{ \frac{s+i\varepsilon_{2 \downarrow}+g}{(s-i\varepsilon)A_{3}^*(s)}\right\}(t){\cal{L}}^{-1}\left\{\frac{A_1(s)}{(s+i\varepsilon)A_3(s)} \right\}(t)\nonumber \\
   &&-i \Delta V_{12}^2 \frac{\Gamma_N}{2\pi} \int_{-\infty}^{+\infty} d\varepsilon [1-f_{N}(\varepsilon)]{\cal{L}}^{-1}\left\{\frac{A_2^*(s)}{(s+i\varepsilon)A_3^*(s)} \right\}(t){\cal{L}}^{-1}\left\{\frac{s+i\varepsilon_{2 \uparrow}+g}{(s-i \varepsilon)A_3(s)} \right\}(t) ,
   \label{chi11t}
\end{eqnarray}
\begin{eqnarray}
\langle \hat c_{2 \downarrow}(t) \hat c_{2 \uparrow}(t) \rangle&=& -n_{1 \uparrow}(0)i\Delta V_{12}^2 {\cal{L}}^{-1}\left\{ \frac{s-i\varepsilon_{2 \uparrow}+g}{A_3^*(s)} \right\}(t){\cal{L}}^{-1}\left\{\frac{A_1(s)}{A_3(s)} \right\}(t) \nonumber \\
&& +n_{2 \uparrow}(0) i\Delta V_{12}^2 {\cal{L}}^{-1}\left\{\frac{1}{A_{3}^*} \right\}(t){\cal{L}}^{-1}\left\{ \frac{A_3(s)-V_{12}^2A_{1}(s)}{(s+i\varepsilon_{2 \uparrow}+g)A_3(s)}\right\}(t) \nonumber \\
&&+[1-n_{1 \downarrow }(0)]i\Delta V_{12}^2{\cal{L}}^{-1}\left\{ \frac{A_2^*(s)}{A_3^*(s)} \right\}(t){\cal{L}}^{-1}\left\{ \frac{s-i\varepsilon_{2 \downarrow}+g}{A_3(s)}\right\}(t) \nonumber \\
&&-[1-n_{2 \downarrow}(0)]i\Delta V_{12}^2{\cal{L}}^{-1}\left\{\frac{A_3^*(s)-V_{12}^2 A_2^*(s)}{(s+i\varepsilon_{2 \downarrow}+g)A_{3}^*(s)} \right\}(t){\cal{L}}^{-1}\left\{ \frac{1}{A_{3}(s)}\right\}(t) \nonumber \\
&& -i\frac{\Gamma_N}{2 \pi} \Delta V_{12}^2 \int_{-\infty}^{+\infty}d\varepsilon [1-f_{N}(\varepsilon)]{\cal{L}}^{-1}\left\{\frac{A^*_3(s)-V_{12}^2A_2^*(s)}{(s+i\varepsilon_{2 \downarrow}+g)(s+i\varepsilon)A^*_3(s)} \right\}(t){\cal{L}}^{-1}\left\{\frac{1}{A_3(s)(s-i\varepsilon)} \right\}(t) \nonumber \\
&&+i\frac{\Gamma_N}{2 \pi} \Delta V_{12}^2 \int_{-\infty}^{+\infty}d\varepsilon f_{N}(\varepsilon) {\cal{L}}^{-1}\left\{\frac{1}{(s-i\varepsilon)A_{3}^*(s)} \right\}(t){\cal{L}}^{-1}\left\{ \frac{A_3(s)-V_{12}^2A_1(s)}{(s+i\varepsilon_{2 \uparrow}+g)(s+i\varepsilon)A_{3}(s)} \right\}(t).
\label{chi22t}
\end{eqnarray}
The interdot current $j_{12\sigma}(t)=2 V_{12}\mbox{\rm Im} \left< \hat{c}_{1\sigma}^{\dagger}(t) \hat{c}_{2\sigma}(t) \right>$ can be expressed for $\sigma=\uparrow$ by the following form
\begin{eqnarray}
j_{12\uparrow}(t) &=& 2V_{12}^{2}
\mbox{Re}  \left\{ n_{1\uparrow}(0) 
{\cal{L}}^{-1}\left\{ \frac{(s-i\varepsilon_{2\uparrow}+g)A_{1}^{*}(s)}{A_{3}^{*}(s)} \right\}(t)
{\cal{L}}^{-1}\left\{ \frac{A_{1}(s)}{A_{3}(s)} \right\}(t)
\nonumber \right. \\
&+& n_{2\uparrow}(0) 
{\cal{L}}^{-1}\left\{ \frac{A_{1}^{*}(s)}{A_{3}^{*}(s)} \right\}(t)
{\cal{L}}^{-1}\left\{ \frac{1-V_{12}^{2}A_{1}(s)/A_{3}(s)}{s+i\varepsilon_{2\uparrow}+g} \right\}(t)
\nonumber  \\
&+& \Delta_{2} [ 1-n_{1\downarrow}(0)] 
{\cal{L}}^{-1}\left\{ \frac{(s-i\varepsilon_{2\uparrow}+g)(s+i\varepsilon_{2\downarrow}+g)}{A_{3}^{*}(s)} \right\}(t)
{\cal{L}}^{-1}\left\{ \frac{(s-i\varepsilon_{2\downarrow}+g)}{A_{3}^{*}(s)} \right\}(t)
\nonumber  \\
&+& \Delta_{2}V_{12}^{2} [ 1-n_{2\downarrow}(0)] 
{\cal{L}}^{-1}\left\{ \frac{s-i\varepsilon_{2\uparrow}+g}{A_{3}^{*}(s)} \right\}(t)
{\cal{L}}^{-1}\left\{ \frac{1}{A_{3}(s)} \right\}(t)
\nonumber  \\
&+&  
\frac{\Delta V_{12}\Gamma_N}{2\pi} \int_{-\infty}^{\infty} d\epsilon [1-f_{N}(\epsilon) ]
{\cal{L}}^{-1}\left\{ \frac{(s-i\varepsilon_{2\downarrow}+g)}{(s+i\epsilon )A_{3}^{*}(s)} \right\}(t)
{\cal{L}}^{-1}\left\{ \frac{1}{(s-i\epsilon)A_{3}(s)} \right\}(t) ,
\nonumber  \\
&-&  \left. 
\frac{\Gamma_N}{2\pi} \int_{-\infty}^{\infty} d\epsilon f_{N}(\epsilon) 
{\cal{L}}^{-1}\left\{ \frac{A_{1}^{*}(s)}{(s+i\epsilon )A_{3}^{*}(s)} \right\}(t)
{\cal{L}}^{-1}\left\{ \frac{v_{12}^{2}A_{1}(s)/A_{3}(s)-1}{(s+i\epsilon)(s+i\varepsilon_{2\uparrow}+g))} \right\}(t)
\right\} ,
\label{eqn.j12}
\end{eqnarray}
the same approach can be used to determine the time-dependent current flowing between superconductor and QD$_1$. The charge flow of spin-$\sigma$ component is given by the standard relation
\begin{eqnarray}
j_{S\sigma}(t)= \frac{d}{dt} \sum_{\bf q} \left<  \hat{c}^{\dagger}_{S{\bf q}\sigma}(t) \hat{c}_{S{\bf q}\sigma}(t)\right>=
2\mbox{\rm Im} \sum_{\bf q} V_{S{\bf q}} \left< \hat{c}_{1\sigma}^{\dagger}(t) \hat{c}_{S{\bf q}\sigma}(t) \right> .
\end{eqnarray}
To calculate the expectation value $\left< \hat{c}_{1\sigma}^{\dagger}(t) \hat{c}_{S{\bf q}\sigma}(t) \right>$ we have to treat the superconducting lead  within the BCS approach (\ref{eq:1}) with the finite pairing gap, and next impose the limit $\Delta_{sc} \rightarrow \infty$. We have previously derived explicit form of $j_{S\sigma}(t)$ for the case  $\varepsilon_{1\sigma}=0$, see Appendix A.4 in Ref.\ \cite{Taranko-2018}. Here we present its generalized version, valid for S-QD$_1$-QD$_2$-N system
\begin{eqnarray}
j_{S\sigma}(t) &=& \frac{\Gamma_{S}^{2}}{2}
\mbox{Re}  \left\{ [ 1-n_{1\sigma}(0)-n_{1-\sigma}(0)] 
{\cal{L}}^{-1}\left\{ \frac{(s+g)^{2}}{w(s)} \right\}(t)
{\cal{L}}^{-1}\left\{ \frac{(s+g)u(s)}{w(s)} \right\}(t)
\nonumber \right. \\
&+& V_{12}^{2} 
[ 1-n_{2\sigma}(0)-n_{2-\sigma}(0)] 
{\cal{L}}^{-1}\left\{ \frac{(s+g)}{w(s)} \right\}(t)
{\cal{L}}^{-1}\left\{ \frac{u(s)}{w(s)} \right\}(t)
\nonumber \\
&+&  \left. 
\frac{\Gamma_N V_{12}^{2}}{2\pi} \int_{-\infty}^{\infty} d\epsilon [1-2f_{N}(\epsilon) ]
{\cal{L}}^{-1}\left\{ \frac{(s+g)}{(s+i\epsilon )w(s)} \right\}(t)
{\cal{L}}^{-1}\left\{ \frac{u(s)}{(s-i\epsilon)w(s)} \right\}(t)
\right\} ,
\label{eqn.A11}
\end{eqnarray}
where $u(s)=s(s+g)+V_{12}^{2}$, and $w(s)=\frac{\Gamma_{S}^{2}}{4}(s+g)^{2}+\left[ s(s+g)+V_{12}^{2}\right]^2$. Let us notice that the currents (\ref{eqn.A11}) are identical for both spins
and we have checked they properly obey the conservation law, Eqn.\ (\ref{eqn.14}). The net current $j_{S}(t)=\sum_{\sigma} j_{S\sigma}(t)$ discussed in Sec.~III is just twice as large as (\ref{eqn.A11}).

\twocolumngrid
\section{Nonequilibrium phenomena}
\label{app_Pairing_chi11}

\begin{figure}
\includegraphics[width=0.92\linewidth]{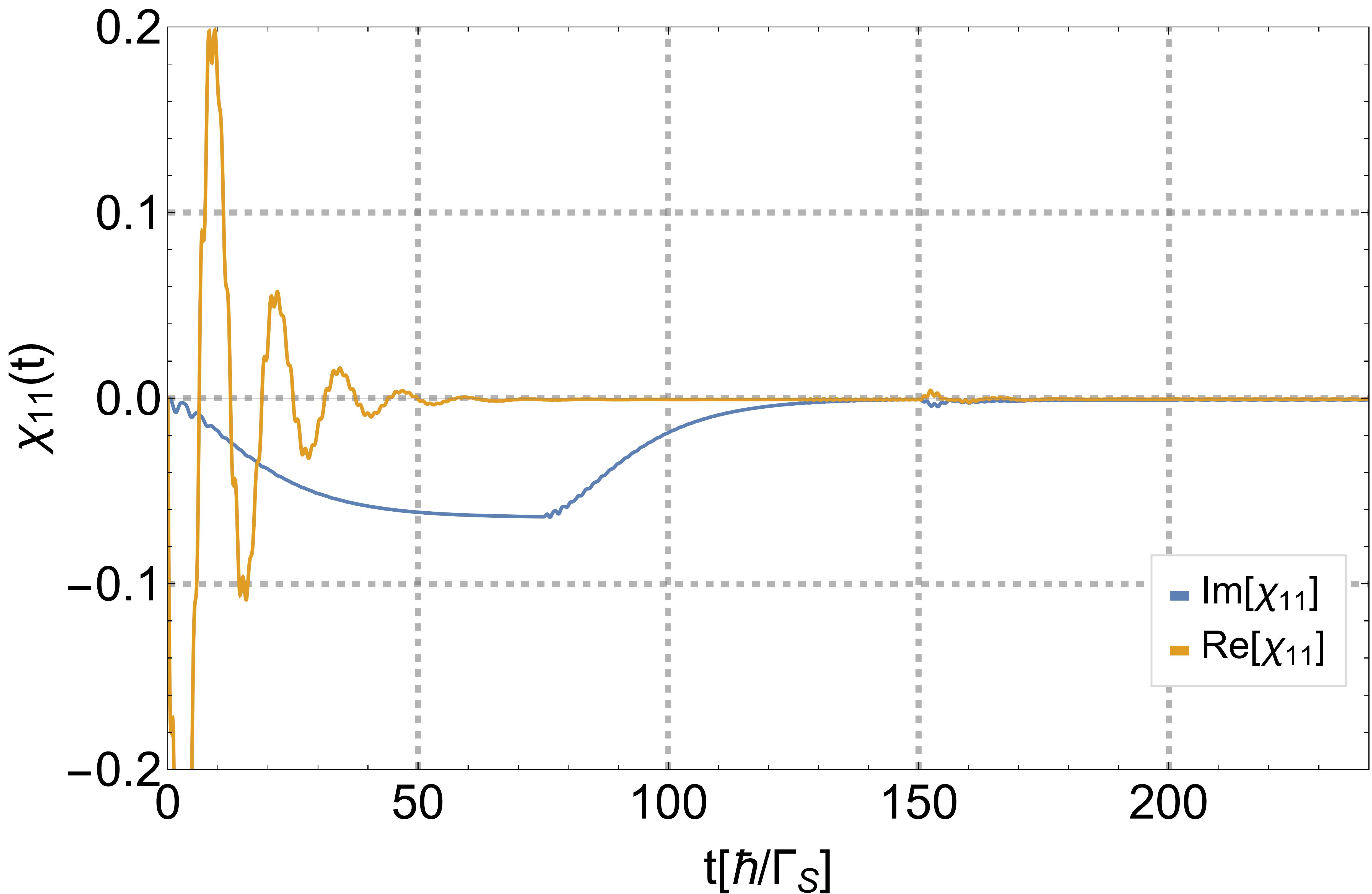}
\includegraphics[width=0.49\linewidth]{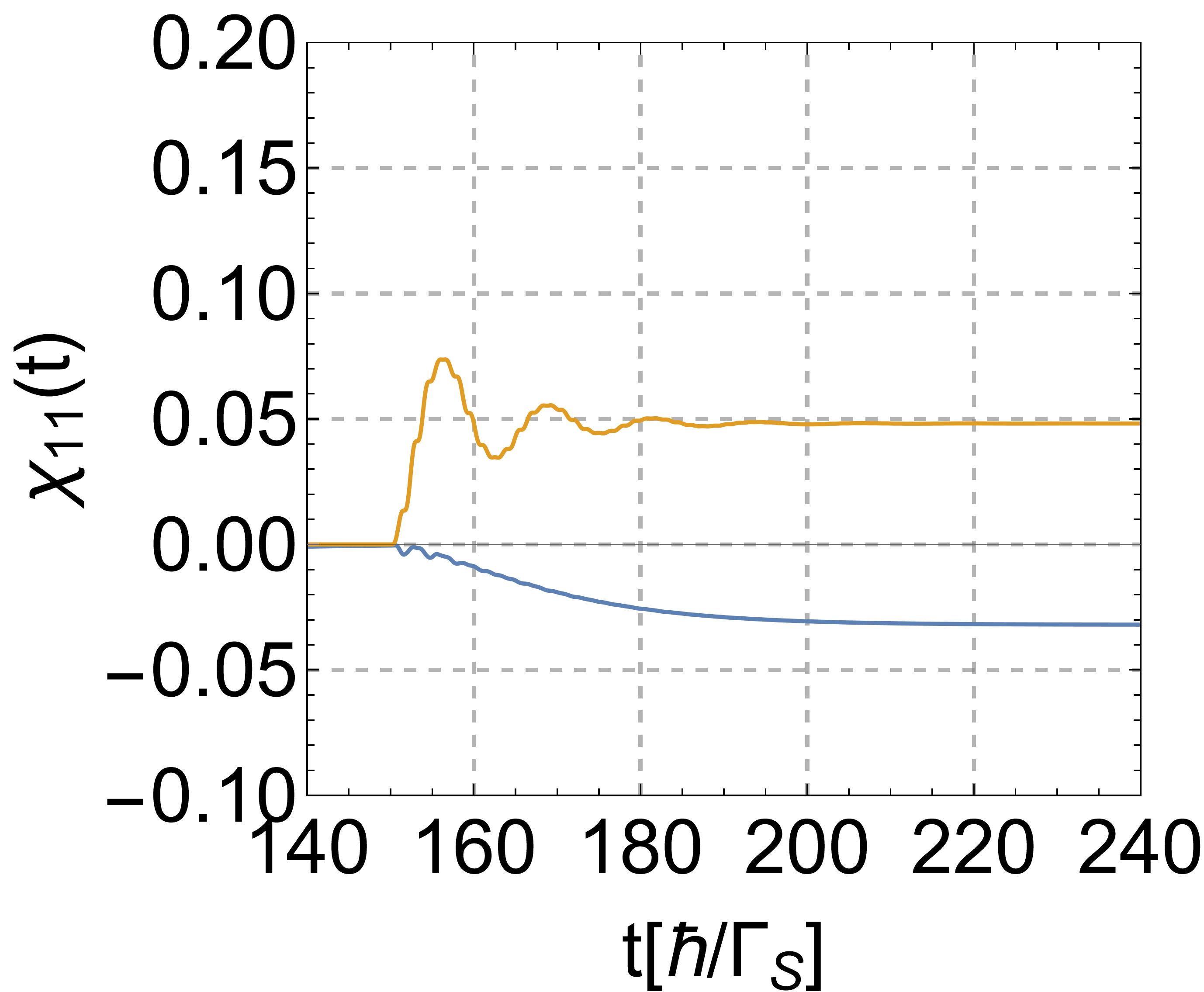}
\includegraphics[width=0.49\linewidth]{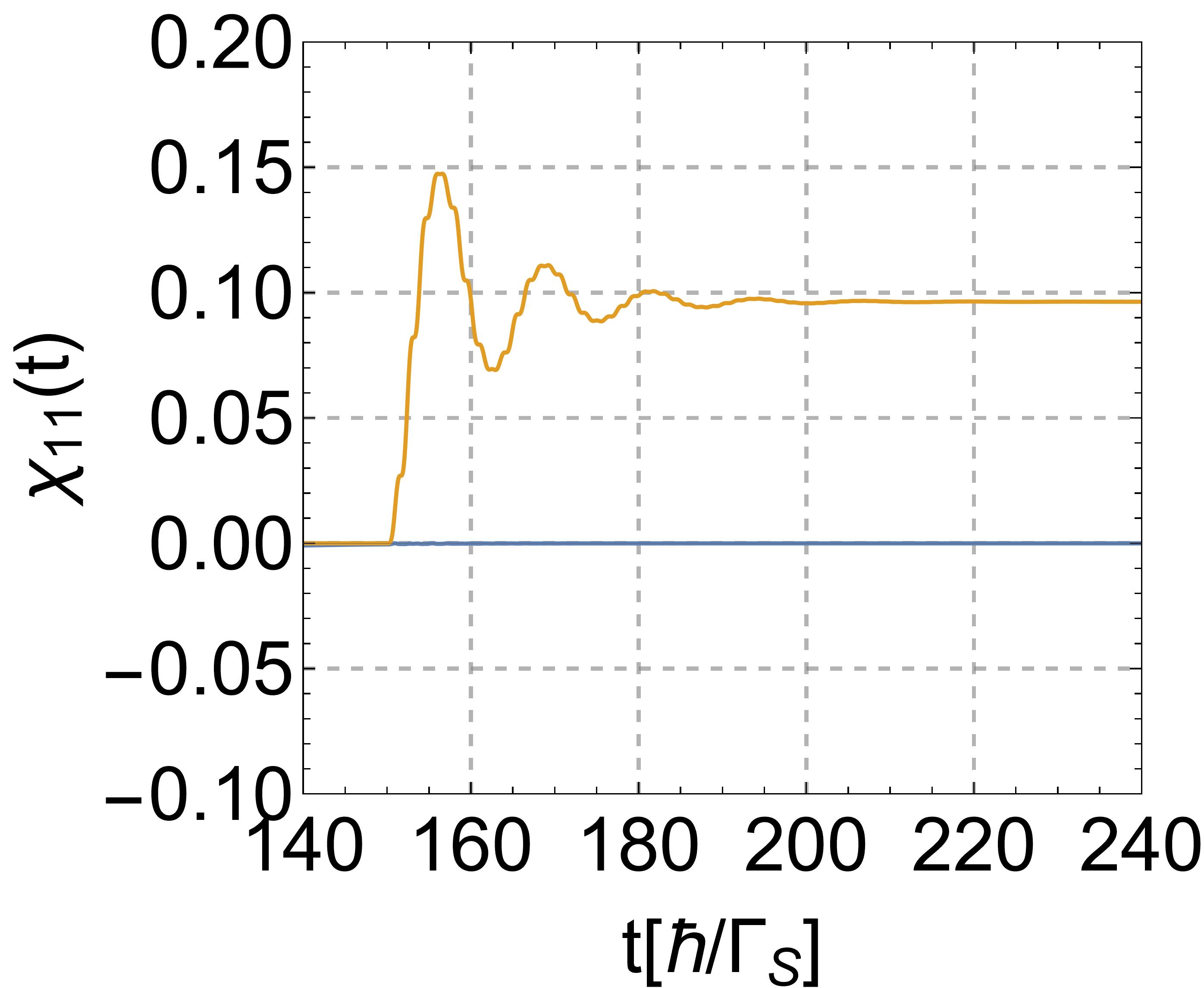}
\caption{The time-dependent pairing $\chi_{11}(t)$ on the quantum dot
attached directly to the superconductor for the same nonequilibrium situation and model parameters as in Fig.~\ref{fig.Zeeman_3pannel}.
}
\label{fig.Apendix_chi11}
\end{figure}

In support for the discussion of the time-dependent processes imposed by the source drain bias (Sec.~\ref{Zeeman.blockade}) we provide here additional
information about the pairing amplitude $\chi_{11}(t)$, the currents and the total charge transmitted between components of our setup.  Figure \ref{fig.Apendix_chi11} shows evolution of the on-dot pairing in the junction biased at $t_1=150 \frac{\hbar}{\Gamma_s}$, applying the voltage $V=2\Gamma_S$ (top panel), $V=5\Gamma_S$ (left-bottom panel) and $V=10\Gamma_S$ (right-bottom panel), respectively. Numerical calculations have been done for the strong inter-dot coupling, $V_{12}=2\Gamma_{S}$, using $\Gamma_{N}=0.2\Gamma_{S}$. 
Depending how large the voltage is, this pairing is partly or completely recovered. We notice that $\chi_{11}(t)$ is in general a complex quantity and its phase differs from the purely real  $\Delta_{sc}$ of  superconductor. This S-QD$_{1}$ segment of our setup can be regarded as the Josephson junction type, where phase difference causes the superflow of electron pairs between QD$_1$ and superconducting lead.

\begin{figure}
\includegraphics[width=0.8\linewidth]{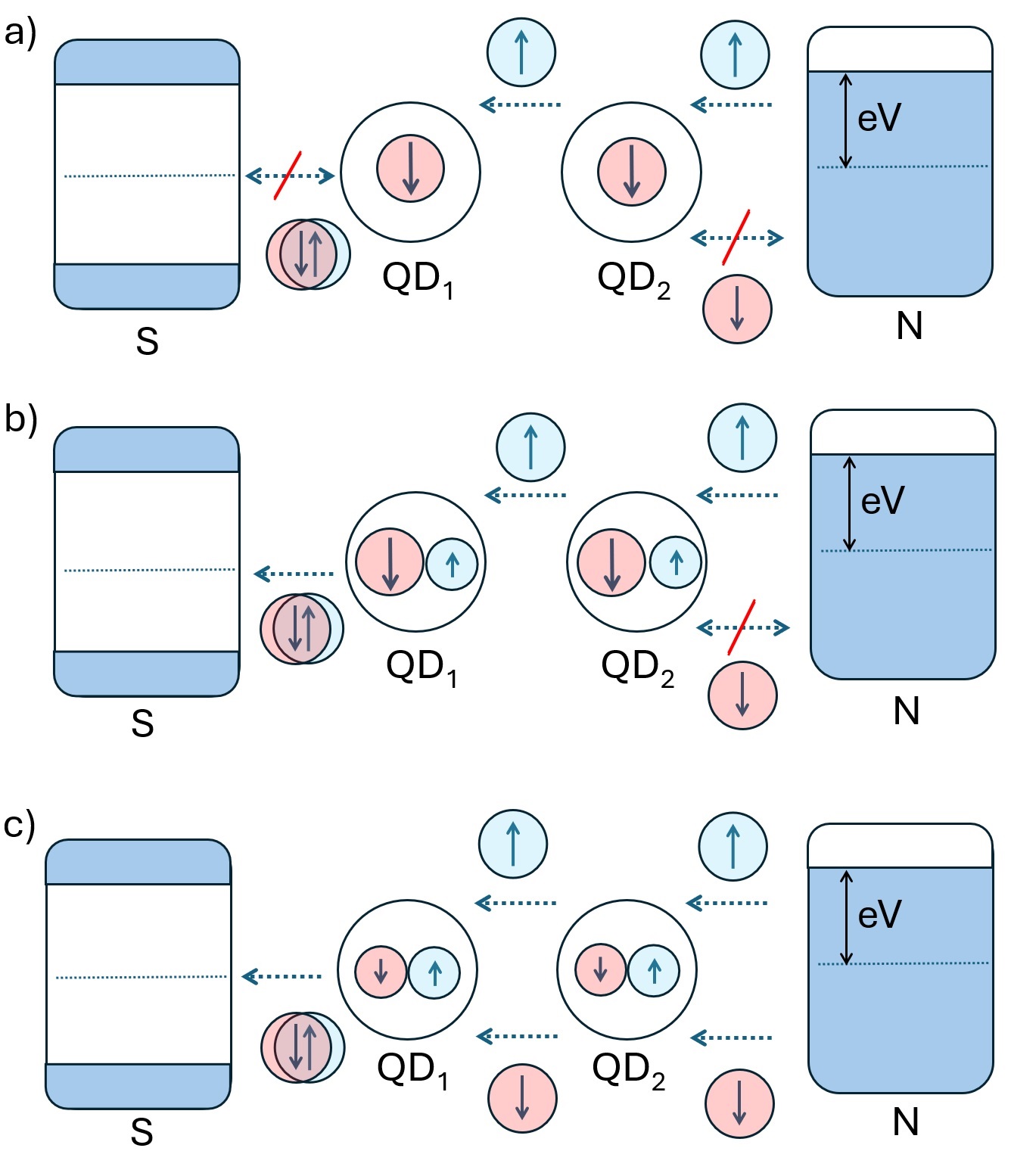}
\caption{Schematic illustration of the subgap charge currents recovered by the strong bias voltage. (a) The system is prepared in the regime where each dot is occupied by single electron of the same spin (here $\downarrow$). Right after applying voltage spin-$\uparrow$ electrons are transferred from N electrode into the system (cf.\ sharp spike of $j_{\uparrow}$ in bottom right panel of Fig.~\ref{fig.Zeeman_current}). (b) Injection of spin-$\uparrow$ electrons to the system allows for creation of local pairs, which can be transferred into S electrode. This process partially depletes spin-$\downarrow$ electrons. (c) The system reaches a steady state with both dots half-filled.}
\label{process}
\end{figure}

In Fig.\ \ref{process} we display schematic explanation how the triplet configuration (imposed by the magnetic field) is overcome by the sufficiently strong bias voltage applied across the junction. Such process occurs via three stages: (i) injection of spin-$\uparrow$ electrons from the normal metallic lead, (ii) appearance of the supercurrent between superconductor and QD$_1$, depleting spin-$\downarrow$ electrons, and finally (iii) a buildup of the steady-state current.

\begin{figure}
\includegraphics[width=0.995\linewidth]{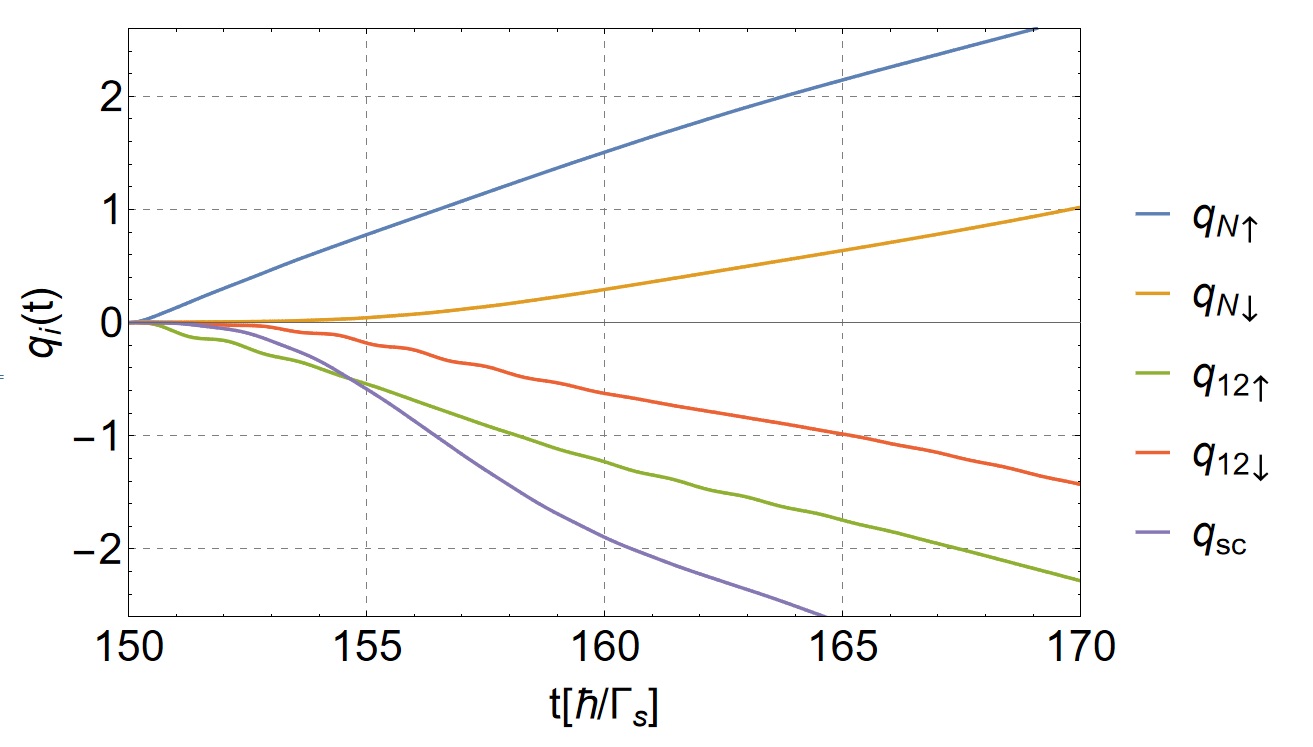}
\caption{The total charge transferred from the metallic lead to QD$_{2}$ after the second quench, $q_{N \sigma}$, the inter-dot charge transfer, $q_{12 \sigma}$, and the superconducting charge flow, $q_{sc}$, obtained for the same model parameters as in Fig.~\ref{fig.Zeeman_current}.}
\label{qj}
\end{figure}

To elucidate the time-dependent currents presented in Fig.~\ref{fig.Zeeman_current}, we show in Fig.~\ref{qj} the charge transferred between individual components of our system. We have computed them numerically from the following integrals
\begin{eqnarray}
    q_{N \sigma}(t)&=&\int_{150}^{t}j_{N\sigma}(t')dt' \;, \label{eq_17} \\
      q_{12 \sigma}(t)&=&\int_{150}^{t}j_{12\sigma}(t')dt' \;, \label{eq_18} \\
        q_{sc}(t)&=&\int_{150}^{t}j_{S}(t')dt' \label{eq_19} \;.
\end{eqnarray}
We observe that after biasing the system, only the spin-$\uparrow$ electrons are transferred from the metal to QD$_2$ (see the blue line in Fig.~\ref{qj}) and then, through the inter-dot current $j_{12 \uparrow}$, further to QD$_1$ (green line). When spin-$\uparrow$ electron arrives at QD$_1$, the local pair can be formed and soon transferred to the superconductor, inducing $q_{sc}$ (purple line). As $q_{sc}$ involves both spin components, this process leaves some spin-$\downarrow$ vacancies on QD$_1$, which can be filled via the inter-dot transfer $q_{12 \downarrow}$ (red line). This, in turn, allows for spin-$\downarrow$ electron to be transferred from the metallic lead to QD$_2$. After a few consecutive cycles (approximately after $t \approx 30 \hbar / \Gamma_S$), these charge transfers (\ref{eq_17}-\ref{eq_19}) linearly increase in time, indicating an onset of the steady-state current.

\section{Dynamics of correlated setup}
\label{correlation_effects}

In the following, we briefly address the correlated setup, $U_j\neq 0$. Competition between the Coulomb interactions $U_j\hat{n}_{j\uparrow}\hat{n}_{j\downarrow}$ and the superconducting proximity effect has been previously studied in detail by several groups under static (see Refs.\ \cite{Bauer-2007,Rodero-2011}) and nonequilibrium conditions \cite{LevyYeyati-2021,Sothmann-2021,Morr-2022,Wrzesniewski-2022,Wegewijs-2023}. In the stationary limit and in the absence of magnetic field, the proximitized quantum dot would be either singly occupied or in the BCS-type state, representing a coherent superposition of the empty and doubly occupied configurations. Their realizations depend on the ratio of $\Gamma_{S}/U_{1}$ and the position of the quantum dot energy levels $\varepsilon_{j\sigma}$ \cite{Bauer-2007,Rodero-2011}.
Here, we investigate the double dot system where initially $QD_{1}$ is in the BCS configuration and we study the evolution after imposing an external magnetic field.

\begin{figure} 
\includegraphics[width=1\columnwidth]{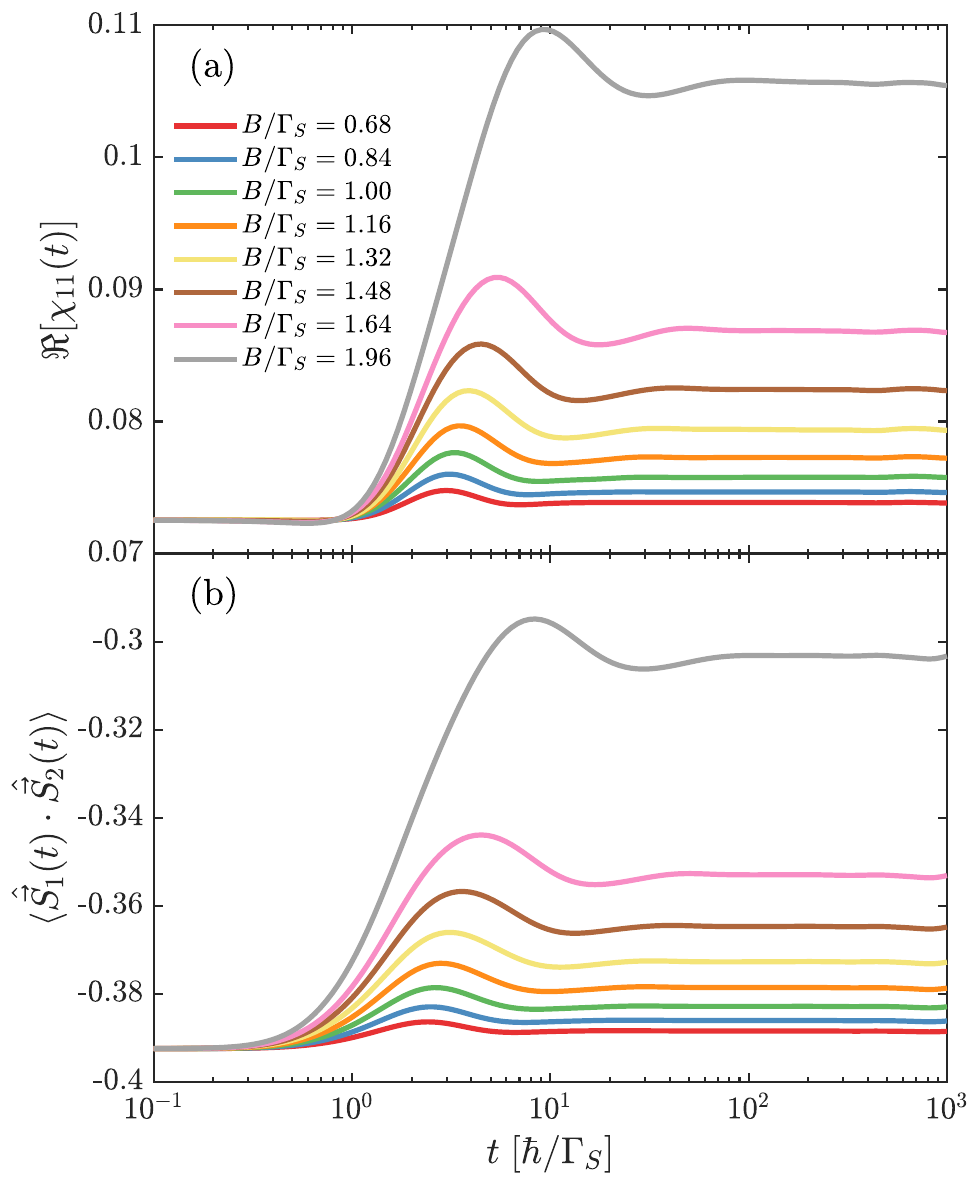}
\caption{(a) Evolution of the electron pairing $\Re[{\chi_{11}(t)] = \langle \hat{c}_{1\downarrow}(t)\hat{c}_{1\uparrow}(t)\rangle}$ and
(b) the spin-spin correlation function $\langle \hat{\vec{S}}_{1}(t)\cdot\hat{\vec{S}}_{2}(t)\rangle$ induced by applying external magnetic field. Numerical results are obtained by tdNRG for the half-filled quantum dots, using the model parameters: \TD{$\mu_{N}=0$,} $U_{1}=U_{2}=0.91$, \TD{$\varepsilon_{j}=-U_{j}/2$,} $\Gamma_N=0.23$, $V_{12}=1.36$ [in units of $\Gamma_S$]. For numerical computations we assumed $\Gamma_S/D=0.22$, where $D$ stands for a half-bandwidth.} 
\label{tdNRG:strong}
\end{figure}

For reliable description of the correlation effects,
we resort to the numerical renormalization group (NRG) method ~\cite{Wilson1975,Bulla2008,NRG_code}.
The main idea of the NRG approach is a logarithmic discretization
of the conduction band, which allows one to map the Hamiltonian to a chain-like form.
The Hamiltonian of such a model is then diagonalized iteratively,
keeping an appropriate number of the low-energy eigenstates.
This method has been successfully used to analyze the stationary
properties of correlated quantum dots coupled to a superconductor \cite{Bauer-2007,Domanski-2016,Wrzesniewski2017Nov,Wojcik2019Jan}.
To account for the dynamical effects, we make use of its time-dependent extension (tdNRG) \cite{Anders2005,Costi2014,WrzesniewskiWeymann-2019}.
Such approach goes beyond the framework of perturbative approximations
and allows for accurate determination of dynamical behavior
in a fully nonperturbative fashion.
Our NRG calculations were carried out with
a discretization parameter $\Lambda=1.8$,
keeping $N_\text{kept}=2000$ states for the Wilson chain length $N=80$.


\begin{figure}
\includegraphics[width=1\columnwidth]{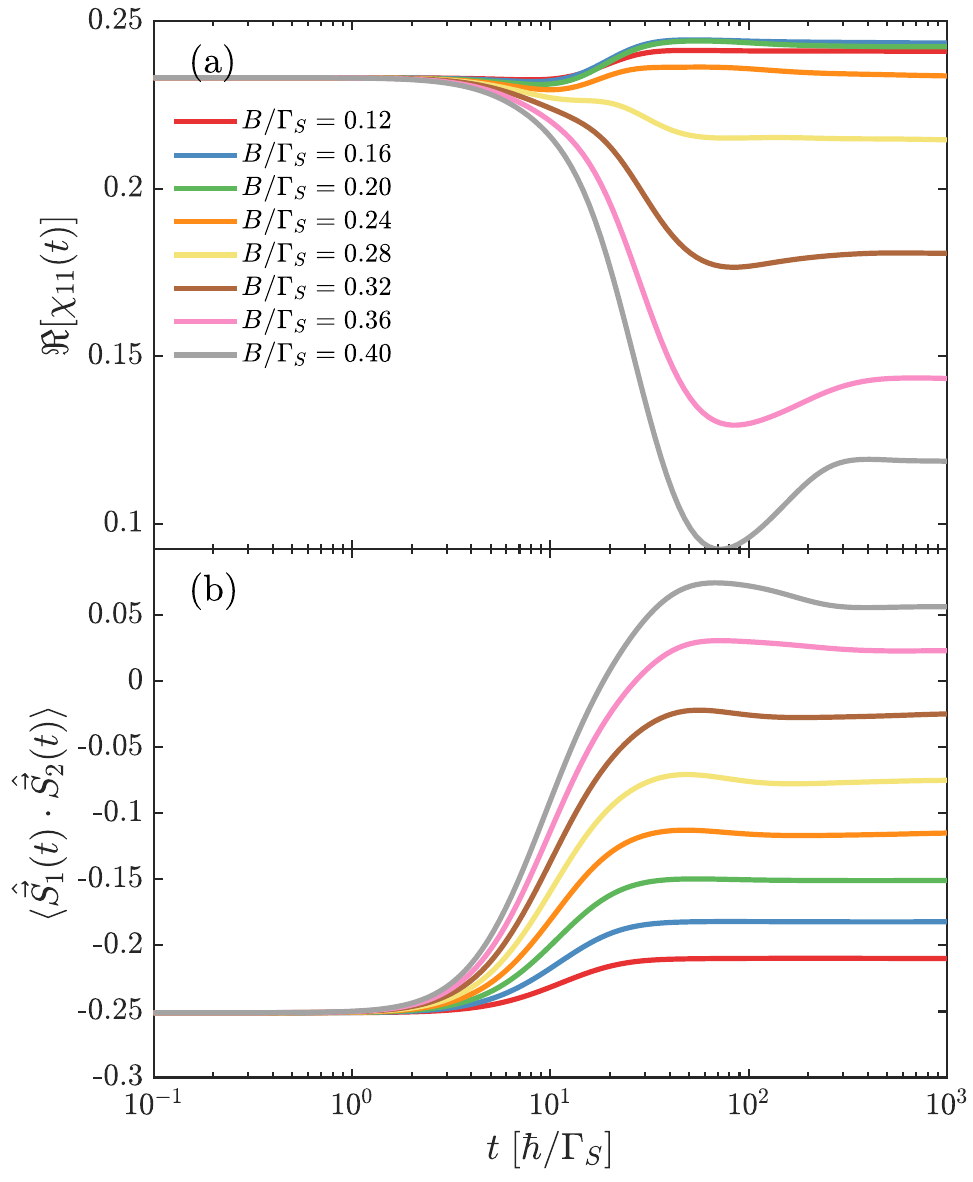}
\caption{The same as in Fig.~\ref{tdNRG:strong}
calculated in the weak inter-dot coupling, 
$V_{12}/\Gamma_{S}=0.045$.}
\label{tdNRG:weak}
\end{figure}

Let us first discuss the case when the coupling between the quantum dots
is strong ($V_{12}>\Gamma_S$). Fig.~\ref{tdNRG:strong} displays
the evolution of the electron pairing on QD$_{1}$, $\Re[\chi_{11}(t)]$,
and the spin-spin correlation function, $\langle \hat{\vec{S}}_{1}(t)\cdot\hat{\vec{S}}_{2}(t)\rangle$,
induced in our setup by external magnetic field at $t=0$. 
First of all, we note that the values of observables
at $t=0$ are strongly dependent on the inter-dot coupling. 
The larger the coupling is, the stronger are the antiferromagnetic
correlations between the quantum dots, and the smaller the pairing becomes.
Since Fig.~~\ref{tdNRG:strong} is calculated for the case of
strong coupling $V_{12}$, one can see that at initial times $\Re[\chi_{11}(t)]$ is very fragile.
Considerable value of $V_{12}$ is needed for extending the electron pairing on both quantum dots, despite the fact that QD$_{2}$ is not directly coupled to the superconductor. Upon switching on the magnetic field, we observe a weak increase
in the electron pairing on QD$_1$. Its formation starts around $t\sim 5 \hbar/\Gamma_S$ and stabilizes for $t\sim 10 \hbar/\Gamma_S$.
This counterintuitive behavior stems from the fact that
the magnetic field suppresses efficiency of the inter-dot hopping, which otherwise would balance superconducting correlations between both quantum dots.
Once this transfer channel is weakened, the pairing is effectively “pushed back” onto QD$_1$,
leading to an increase of $\Re[{\chi_{11}(t\rightarrow \infty)}]$.
Moreover, the initial strong antiferromagnetic spin-spin correlations
are only partially reduced by the applied quench,
even for significant values of the magnetic field, see Fig.~\ref{tdNRG:strong}(b).
Here, we note that time dependencies of both evaluated quantities
reveal similar corresponding dynamics, and the magnetic field
dependence of the long-time behavior of both $\Re[\chi_{11}(t)]$
and $\langle \hat{\vec{S}}_{1}(t)\cdot\hat{\vec{S}}_{2}(t)\rangle$ is monotonic.


Figure \ref{tdNRG:weak} presents evolution of the on-dot pairing
and the spin-spin correlations between the quantum dots
in the case of weak inter-dot coupling. Here, at initial times,
we find significantly higher electron pairing  accumulated on QD$_1$,
in comparison to the case shown in Fig.~\ref{tdNRG:strong}.
Consequently, the absolute value of antiferromagnetic spin-spin correlations is much weaker.

After sudden switching of the magnetic field $B$, the time-dependent evolution reveals qualitatively distinct
behavior of the local pairing on QD${_1}$ and the spin–spin correlations between the two dots.
Influence of $B$ on the spin-spin correlations is straightforward:
regardless of its magnitude, $\langle \hat{\vec{S}}_{1}(t)\cdot\hat{\vec{S}}_{2}(t)\rangle$
evolves from negative toward positive values, signaling
a transition from the singlet to triplet configuration.
For sufficiently strong $B$, the system eventually stabilizes in a robust triplet state 
$\langle \hat{\vec{S}}_{1}(t)\cdot\hat{\vec{S}}_{2}(t)\rangle>0$ associated with the subgap blockade.

Dynamics of the electron pairing in the proximitized quantum dot is, however, more subtle. For the weak magnetic fields ($B\lesssim V_{12}$), 
$\Re[\chi_{11}(t)]$ is actually enhanced compared to the zero-field case.
For the stronger fields ($B\gtrsim V_{12}$),
the Zeeman splitting dominates therefore the local pairing  becomes suppressed.
Consequently, while the asymptotic (long-time) spin-spin correlations
show a monotonic increase with $B$, the steady-state pairing
amplitude on QD$_1$ exhibits a non-monotonic dependence,
first rising for small values of $B$ and then decreasing at larger fields. This behavior is caused by rearrangement of the bound states
energies due to the magnetic field.

\bibliography{biblio_blockade}

\end{document}